\documentclass{aa}
\usepackage{graphicx}
\usepackage{txfonts}
\usepackage{epsf}  
\usepackage{amssymb}  
\usepackage{rotating}
\usepackage{lscape} 
\usepackage{longtable}
\usepackage{natbib}
 \usepackage{hyperref}
\bibpunct{(}{)}{;}{a}{}{,} % to follow the A&A style
\begin{document}
%
%
% Abbreviations:
 \newcommand{\kms}{km\,s$^{-1}$}
 \newcommand{\teff}{$T_{\rm eff}$}
 \newcommand{\teffii}{T_\mathrm{eff}}
 \newcommand{\logg}{$\log g$}
 \newcommand{\feh}{[Fe/H]}
 \newcommand{\msun}{${\rm M}_{\odot}$}
\newcommand{\percent}{\,{\%}}
\newcommand{\kepler}{{\em Kepler}}

% HB, new commands
\newcommand{\vsini}{$v \sin i$}
\newcommand{\feone}{Fe\,{\sc I}}
\newcommand{\fetwo}{Fe\,{\sc II}}

\title{Atmospheric parameters of 82 red giants in the {\em Kepler} field.
\thanks{Based on observations made with the Nordic Optical Telescope, operated
on the island of La Palma jointly by Denmark, Finland, Iceland,
Norway, and Sweden, in the Spanish Observatorio del Roque de los
Muchachos of the Instituto de Astrofisica de Canarias. Reduced spectra are only available in electronic form
at the CDS via anonymous ftp to cdsarc.u-strasbg.fr (130.79.128.5)
or via http://cdsweb.u-strasbg.fr/cgi-bin/qcat?J/A+A/}}

\titlerunning{Atmospheric parameters of red giants}
\authorrunning{A. O. Thygesen et al.}
\author{
A. O.~Thygesen\inst{1,2} % organiser, writer of proposals, data reduction, paper writer
\and
S.~Frandsen\inst{1}
\and
H.~Bruntt\inst{1}
\and
T.~Kallinger\inst{3}
\and
M. F.~Andersen\inst{1}
\and
Y. P.~Elsworth\inst{4}
\and
S.~Hekker\inst{5,4}
\and
C.~Karoff\inst{1}
\and
D.~Stello\inst{6}
\and
K.~Brogaard\inst{7}
\and
C.~Burke\inst{8}
\and
D. A.~Caldwell\inst{8}
\and
J. L.~Christiansen\inst{8}
} 
\offprints{A. O.~Thygesen}
\mail{aot06@phys.au.dk}
\institute{Department of Physics and Astronomy, Aarhus University, DK-8000 Aarhus C, Denmark.
\and Zentrum f\"{u}r Astronomie der Universit\"{a}t Heidelberg, Landessternwarte, K\"{o}nigstuhl 12, 69117 Heidelberg, Germany.
\and Instituut voor Sterrenkunde, K.U. Leuven, Celestijnenlaan 200D, 3001 Leuven, Belgium.
\and School of Physics and Astronomy, University of Birmingham, Edgbaston, Birmingham B15 2TT, UK.
\and Astronomical Institute, 'Anton Pannekoek', University of Amsterdam, PO Box 94249, 1090 GE Amsterdam, The Netherlands.
\and Sydney Institute for Astronomy, School of Physics, University of Sydney, NSW 2006, Australia.
\and Department of Physics and Astronomy, University of Victoria, PO Box 3055, Victoria, BC V8W 3P6, Canada.
\and SETI Institute/NASA Ames Research Center, Moffett Field, CA 94035.}
\date{Received 19-03 2012 ; Accepted 23-05 2012 }
\abstract
%Context
{Accurate fundamental parameters of stars are essential for the asteroseismic analysis of data from the NASA \kepler\ mission.} 
%Aims
{We aim at determining accurate atmospheric parameters and the abundance pattern for a sample of 82 red giants that are targets for the \kepler\ mission.}
% Methods
{We have used high-resolution, high signal-to-noise spectra from three different spectrographs. We used the iterative spectral synthesis method VWA to derive the fundamental parameters from carefully selected high-quality iron lines. After determination of the fundamental parameters, abundances of 13 elements were measured using equivalent widths of the spectral lines.}
% Results
{We identify discrepancies in \logg\ and \feh, compared to the parameters based on photometric indices in the \kepler\ Input Catalogue (larger than 2.0 dex for \logg\ and \feh\ for individual stars). The \teff\ found from spectroscopy and photometry shows good agreement within the uncertainties. We find good agreement between the spectroscopic \logg\ and the \logg\ derived from asteroseismology. Also, we see indications of a potential metallicity effect on the stellar oscillations.}
% Conclusions
{We have determined the fundamental parameters and element abundances of 82 red giants. The large discrepancies between the spectroscopic \logg\ and \feh\ and values in the \kepler\ Input Catalogue emphasize the need for further detailed spectroscopic follow-up of the \kepler\ targets in order to produce reliable results from the asteroseismic analysis.}
\maketitle
%
%
%
%__________________________________Introduction
\section{Introduction \label{sec:intro}}
Since 2009 the NASA \kepler\ mission has continuously been measuring the flux for thousands of stars within the same field on the sky. Although the main objective of the mission is the detection of extra-solar planets, the high-precision lightcurves are excellent data for asteroseismic investigations. In particular, the study of oscillating red giants has taken a huge leap forward. Before the launch of \kepler\ and CoRoT only a handful of giants were known to oscillate. This number has now increased to more than 10,000 \citep{hekker1}. Already several important results have emerged, most noticeably the ability to distinguish between red giant branch (RGB) and clump giants \citep{bedding1}.

To obtain robust results from the analysis and interpretation of the oscillation data provided by \kepler, accurate fundamental parameters of the targets are needed as previously discussed by \citet{brown, basu, hekker2}. 

The \kepler\ Input Catalogue (KIC, \citealt{brown2}) provides values for the effective temperature, \teff, surface gravity, \logg, and iron abundance, \feh, derived from photometric indices. This has so far been the only source for the fundamental parameters of the target stars. These parameters have previously been shown to be inaccurate by \citet{joanna}, who investigated more than 100 main-sequence and giant stars in the \kepler\ field. Their result was confirmed for an additional small sample of \kepler\ red giants in \citet{bruntt1}, hereafter Paper {\sc I}. Also, \citet{bruntt5} identify serious metallicity discrepancies in a sample of 93 solar-type stars.

In this paper we present results for a large sample of red giants observed with high-resolution spectrographs. We focused the observations on stars with a large range in metallicity, from the bottom to the top of the giant branch including both shell hydrogen-burning and core helium-burning stars. From the spectra we obtain accurate fundamental parameters as well as abundances for several elements. These are needed in order to put strong constraints on the stellar models. We also tried to find stellar {\it twins} with similar asteroseismic parameters, but different metallicities.

\section{Target selection}
The targets were selected using KIC and the asteroseismic data available from the \kepler\ satellite. We used the large frequency separation, $\Delta\nu$, which is proportional to the mean density, $\overline\rho$, of the star, ($\Delta\nu \propto \overline\rho^{1/2}$), and the frequency of maximum oscillation power, $\nu_\mathrm{max}$, determined from the power spectra of the lightcurves, when these parameters were known at the time of the selection. 

In this selection process, we were hampered by the fact that we had to use effective temperatures from the KIC. However, as shown in Paper {\sc I} the effective temperatures found in the KIC are in reasonable agreement with the spectroscopic values, at least for targets with \feh\ $>-1.0$ dex, so to first order the identification should be reliable. Also, it was difficult to get values for stars with a low $\Delta\nu$, since only observations from quarters Q0-2 of the \kepler\ mission were used for the stellar {\it twin} identification, resulting in few low-$\Delta\nu$ twins. These quarters had durations of 10 days (Q0), a short quarter  of 34 days (Q1) and one full quarter of 90 days of observations (Q2). Most of our targets have been observed by \kepler\ in it's long cadence mode (29.4 min integration times). This is short enough to properly sample the oscillations in the giants as they oscillate at very low frequencies, well below the Nyquist frequency for the long cadence observations ($\nu_{Ny}=283~\mu \mathrm{Hz}$). We also selected a few sub-giants which oscillate at higher frequencies and hence require short cadence observations ($\sim$1 min integration times).

A number of targets were also chosen entirely because they had extreme metallicities according to the KIC (\feh $<-0.8$ and \feh $>0.3$).

All targets were chosen to have a V-magnitude $<12$ in order to get the desired signal-to-noise (S/N $\geq80$) in the spectra for the spectroscopic analysis. 

\section{Observations}
Three different telescopes equipped with high-resolution echelle spectrographs were used to obtain the data. On the 2.56-m Nordic Optical Telescope (NOT) we used the FIbre-fed Echelle Spectrograph (FIES), covering the spectral range from 370-730 nm, using the high-resolution mode ($R=67.000$). We also aquired data from the 3.6-m Canadian-French-Hawaiian Telescope (CFHT) and 2.0-m Telescope Bernard Lyot (TBL), using the ESPaDOnS (CFHT) and NARVAL (TBL) spectrographs. These instruments provide a more extended wavelength coverage than FIES, covering the range 370-1050 nm with resolutions $R=80.000$ (ESPaDOnS) and $R=75.000$ (NARVAL).

The observations using CFHT and TBL were both carried out in service mode from May to September 2010, as a part of two larger programs for \kepler\ follow-up. The targets observed with CFHT and TBL were among the brightest giants in the sample and had exposure times $\leq15$ min. The NOT observations were performed in July 2010. Exposure times varied according to the brightness of the targets, from a few minutes to 1.5 hours. Could the desired S/N not be reached within a 30 minute exposure, the observations were split into multiple exposures of a maximum length of 30 minutes each. This was done to avoid problems with cosmic rays hitting the CCD. The individual spectra were merged after the data reduction to increase the S/N.

Calibration images for all spectrographs were obtained as standard procedure in the beginning of each observing night. They consist of 1 Thorium-Argon (ThAr), 7 bias and 21 halogen flat exposures in the case of FIES and 1 ThAr, 3 bias and 40 flats for NARVAL and ESPaDOnS. Further, all science exposures using FIES were bracketed with a ThAr exposure to ensure accurate wavelength calibration in the reduction process. The observations and instruments used are summarized in Table~\ref{obstable}.

The reduction of the CFHT and TBL observations were performed as part of the service program, using the ESpRIT pipeline \citep{donati}. The FIEStool\footnote{http://www.not.iac.es/instruments/fies/fiestool/FIEStool.html} pipeline was used for the reduction of the NOT observations. Both pipelines perform the standard tasks involved in the reduction of echelle spectra; bias substraction, order identification, substraction of scattered light, flat fielding, wavelength calibration and extraction of the spectral orders.

\begin{table*}%
\centering
\caption{Summary of the observations and instruments used.}
\begin{tabular}{llllcrrl}
	\hline\hline
	Telescope & Aperture & Instrument & CCD & Resolution & Coverage & No. of targets & S/N \\
	\hline
	NOT & 2.56m & FIES & E2V 2k $\times$ 2k & 67.000 & 370$-$730 nm & 62 & 80-100\\
	CFHT & 3.60m & ESPaDOnS & E2V 2k $\times$ 4.5k & 80.000 & 370$-$1050 nm & 15 & $\approx$ 200 \\
	TBL & 2.00m  & NARVAL & E2V 2k $\times$ 4.5k & 75.000 & 370$-$1050 nm & 5 & $\approx$ 200\\
	\hline\hline
\end{tabular}
\label{obstable}
\end{table*}

\section{Spectroscopic analysis}
The high-resolution spectra were used to determine \teff, \logg, microturbulent velocity ($\xi_t$) and abundances for several elements. We used the semi-automatic software package, VWA\footnote{Available from https://sites.google.com/site/vikingpowersoftware/} \citep{bruntt2,bruntt3,bruntt4} for the analysis. Neutral and singly-ionized iron lines (\feone\ and \fetwo) in the wavelength range 4500Å to 7000Å for the NOT and 4000Å to 8800Å for CFHT and TBL were used to derive the atmospheric parameters. This was done by removing any correlation between the abundances of \feone, excitation potential (EP), and equivalent width (EW) of the spectral lines. In the initial analysis, agreement between the \feone\ and \fetwo\ abundance was required in order to determine \logg. The method is described in greater detail in Paper {\sc I}.

In general we used lines with $0<$EW$<150$ mÅ as the weak lines are the most sensitive to changes in the fundamental parameters. For the brightest of our targets a larger number of lines were of good quality due to the high S/N so here the analysis was restricted to lines with EW$<90$ mÅ. 

It is essential to use as many of the same lines as possible for every star, because the value of the fundamental parameters derived from EW measurements can be sensitive to the particular choice of lines. We therefore compiled a list of lines present in at least 70\percent\ of the targets analysed in Paper {\sc I} and used this as input for the analysis presented here. However, since we deliberately selected a wide range in metallicities, one would expect large star-to-star differences in the number of lines that could be used for the analysis. All spectra were therefore inspected in detail and additional lines added when appropriate. This mainly consisted of non-blended iron lines, as these are preferred for the fundamental parameter determination, but also non-blended lines of other elements were selected (around 30-40 lines per star). In a typical analysis some 400 lines of different elements were used, of which around 130 were iron lines. The iron lines were used to determine the fundamental parameters. The remaining lines were only used to derive abundances.

\subsection{The new oscillator strength correction}
When analysing the solar spectrum using MARCS atmosphere models \citep{gustafsson} as done in this work, we found that there is a positive correlation between the iron abundances and the EP, suggesting a higher \teff\ for the solar atmosphere. However, as the solar \teff\ $=5777\pm10$ K is well determined \citep{smalley}, this indicates a problem with the abundances derived from the lines. This is most likely due to a problem with either the oscillator strengths of the lines (log-$gf$) or with the assumptions in the 1D LTE models used (see \citealt{lobel} for a discussion of corrections). The log-$gf$ values used here are all taken from the Vienna Atomic Line Database (VALD, \citealt{kupka}). To remedy the apparent problem with the oscillator strengths, a list of corrections to the log-$gf$ values for a large number of lines has been calculated.

We found that using only lines for which a log-$gf$ correction exists, greatly reduced the scatter of the abundances when analysing stars other than the Sun (see \citealt{bruntt3} for an illustration), even when there were significant differences in the thermal structure of the line formation regions, compared to the Sun. Here we use a new set of corrections, calculated for more than 1200 lines \citep{bruntt5}, based on the Fourier Transform Spectrometer Kitt Peak Solar Flux Atlas. This greatly increased the number of useful lines, especially lines with low EP, compared to the analysis in Paper {\sc I}.

Before applying these corrections to our targets, they were tested on a spectrum of the Sun. We used a solar spectrum obtained with FIES from a day-time blue-sky exposure. The reduced spectrum was treated in exactly the same way as all other targets, albeit with a much higher S/N. After applying the log-$gf$ corrections, the analysis yielded \teff\ $=5777\pm20$ K, \logg\ $=4.45\pm0.03$ dex and \feh\ $=0.00\pm0.05$ dex showing that our corrections are reliable. We used 220 iron lines for the analysis of the solar spectrum. The uncertainties quoted are the internal precision in VWA, and should not be taken as absolute uncertainties that will indeed be larger. 

One small disadvantage of only using log-$gf$-corrected lines is that only lines that are in common between the giants and the Sun are used in the analysis. As giants display a large number of lines, many of which are being well-fitted by the software, a number of otherwise well-suited lines are therefore being discarded as they are not present in the Sun. But this is compensated by the reduction of both the uncertainties and scatter of the abundances derived from the individual lines. 

For consistency we re-analysed the targets that were presented in Paper {\sc I}, with our new set of log-$gf$ corrections. This resulted in slightly different values for the parameters (\teff, \logg, \feh) as well as a reduction of their uncertainties.

\subsection{Determining macroturbulent and rotational velocity}
In order to determine the macroturbulent velocity and line-of-sight rotational velocity, \vsini, synthetic line profiles were calculated for isolated, weak lines over the entire wavelength range. Line profiles were typically calculated for 20 different regions for each spectrum. The two parameters were adjusted until a satisfactory match between the observed and synthetic line profile was achieved. This was determined by inspecting the residuals by eye. An average of the macroturbulent velocity and \vsini\ found at different wavelength ranges were used as the best estimates. We note that for giants it is hard to distinguish between line broadening from rotation and macroturbulence; hence a more appropriate description would be a total macroscopic broadening, including both effects. But due to the analysis software the two parameters had to be separated.

\section{Results from spectroscopic analysis}
In this section we present the results obtained from the spectroscopic analysis as well as comparisons with the parameters from asteroseismology and the KIC. 

\subsection{Comparison stars}
\begin{figure}%
\includegraphics[width=\columnwidth]{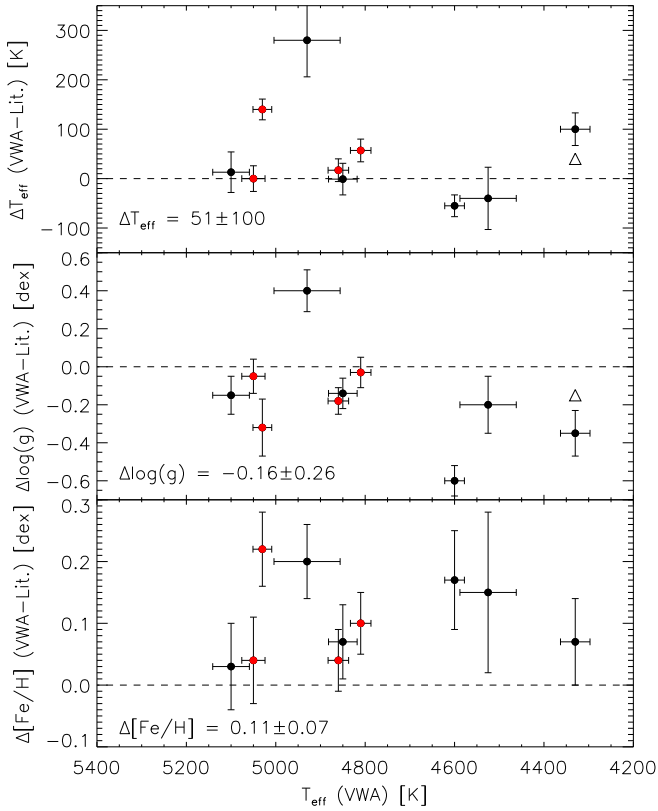}%
\caption{Comparison of the parameter values determined from VWA and literature values for 10 bright giants with well-determined parameters. All results are the re-determined values from this work. The four comparison stars added to the sample from Paper {\sc I} are marked in red. The triangle indicates the asteroseismic values for $\alpha$ Boo.}%
\label{compstarfig}%
\end{figure}

To test the reliability of our analysis, we re-analysed the six bright giants, also reported in Paper {\sc I}, as well as an additional four bright giants. All ten stars have been analysed several times in the literature and are thought to have well-determined parameters. Nine of the stars were chosen from the PASTEL catalogue of \citet{soubiran}, with the last one, HD205512, chosen from \citet{luck}. The four new stars are located in the vicinity of the \kepler\ field. The comparison values were chosen from the most recent analysis that provided values for all three parameters (\teff, \logg, \feh). We only picked results where a spectroscopic analysis had been performed. The result for these stars are presented in Table~\ref{compstars} and the residuals are shown in Fig.~\ref{compstarfig}, with the four new stars shown in red. It is evident from the plots that there is some scatter present between the values from the literature and the values determined by our analysis. One would expect some disagreement between the different methods as indeed observed. That we on average find a larger \feh\ compared to the literature can be explained by the higher \teff\ also found in our analysis, which affects the derived abundances. We use the mean offset as a measure of the 'systematic uncertainty' present in our analysis and add $\Delta$\teff, $\Delta$\logg\ and $\Delta$\feh\ quadratically to the internal uncertainties in VWA. Fundamental parameters for $\alpha$ Boo (Arcturus), which is the coolest star in our comparison sample, are also available from asteroseismology \citep{kallinger2}, who found \teff\ $=4290$ K and \logg\ $=1.45$ dex, showing good agreement with the spectroscopic \teff\ and with a \logg-value roughly half way between our result and the literature value.

\begin{table*}%
\centering
\caption{Values of spectroscopic parameters for 10 bright giants determined from VWA and the literature. Uncertainties on the VWA results are the internal precision only. Uncertainties on the PASTEL values are 80 K on \teff, 0.1 dex on \logg\ and 0.1 dex on \feh. Uncertainties on the Luck \& Heiter result are 100 K on \teff, 0.1 dex on \logg\ and 0.1 dex on \feh.}
\begin{tabular}{l|llr|llr|l}
\hline\hline
   &               & VWA           &              &                &     Lit.        &             & \\
ID & \teff & \logg & \feh & \teff & \logg & \feh & Reference \\
\hline
$\alpha$ Mon &	4850$\pm$32 &	2.60$\pm$0.08	& 0.05$\pm$0.06 &	4851 &	2.74 &	$-$0.02 & PASTEL \\
$\mu$ Leo &	4525$\pm$63 &	2.70$\pm$0.15 &	0.44$\pm$0.13 &	4565 &	2.90 &	0.29 & PASTEL \\
$\alpha$ Boo &	4330$\pm$33 &	1.30$\pm$0.12	&	$-$0.56$\pm$0.07 &	4230 &	1.65 &	$-$0.63 & PASTEL \\
$\mu$ Peg	& 5100$\pm$41 &	2.90$\pm$0.10	&	0.06$\pm$0.07 &	5087 &	3.05 & 0.03 & PASTEL \\
$\psi$ UMa &	4600$\pm$22 &	1.95$\pm$0.08	&	0.03$\pm$0.08	& 4655 &	2.55 &	$-$0.14 & PASTEL \\
$\lambda$ Peg &	4930$\pm$74 &	2.40$\pm$0.11 &	$-$0.06$\pm$0.06 &	4650 &	2.00 &	$-$0.26 & PASTEL \\
HD091190 &	5030$\pm$21 &	2.75$\pm$0.15	&	0.07$\pm$0.06 &	4890 &	3.07 &	$-$0.15 & PASTEL \\
HD186675 &	5050$\pm$26 &	2.80$\pm$0.09	&	0.02$\pm$0.07 &	5050 &	2.85 &	$-$0.02 & PASTEL\\
HD197989 &	4860$\pm$23 &	2.60$\pm$0.07	&	$-$0.07$\pm$0.05 &	4843 &	2.78 &	$-$0.11 & PASTEL\\
HD205512 &	4810$\pm$23 &	2.50$\pm$0.08 &	0.11$\pm$0.05 &	4753 &	2.53 &	0.01 & (Luck \& Heiter 2007) \\
\hline\hline
\end{tabular}
\label{compstars}
\end{table*}

\subsection{Radial velocity measurements}
To identify potential Population {\sc II} targets in our sample we measured the radial velocities (RV) using a cross-correlation (CCF) approach. For each target we calculated the CCF between the observed spectrum and a template spectrum at laboratory wavelengths. The template used for each star was chosen as one of three different templates, corresponding to low-, solar- and high-metallicity targets (\feh\ $=-2.0$, 0.0 and 1.0 dex respectively). Individual CCF's were calculated for each spectral order, spanning the same wavelength range as for the abundance analysis. This method provided 40 CCF's for each observed spectrum that were combined into a total CCF for each target using the maximum-likelihood approach described in detail by \citet{zucker}. The combined CCF was fitted with a Gaussian to derive the RV's. Uncertainties on the measurements were estimated as $\sigma_\mathrm{RV}=(\mathrm{FWHM}/\sqrt{N_\mathrm{ccf}-1})$ where FWHM is the full width at half maximum of the Gaussian and $N_\mathrm{ccf}$ is the number of CCF's used in the combined cross-correlation. Finally we corrected the measurements for barycentric motion. We identify three Pop. {\sc II} stars in our sample. KIC 5698156 and 8017159 we classify as Pop. {\sc II} based on their metallicities and radial velocities (\feh$=-1.33$ dex, $\mathrm{RV}=-381.2$ \kms\ and \feh$=-1.95$ dex, $\mathrm{RV}=-357.4$ \kms\ respectively). KIC 8017159 was already reported as a Pop. {\sc II} star in Paper {\sc I}. KIC 7693833 is identified as a Pop. {\sc II} star, solely based on the very low metallicity we find for this star (\feh\ $=-2.23$ dex).

\subsection{Fundamental parameters from spectroscopy}
The fundamental parameters for all targets in our sample are presented in Table~\ref{atmostable}. We list the KIC values as well as parameters determined from asteroseismology, using either the scaling relations of \citet{white} or the method of \citet{kallinger}. A comparison between the two methods gives a mean difference of 0.03 dex on \logg, which is negligible compared to the typical uncertainty of 0.2 dex from a purely spectroscopic analysis. 

As illustrated in Fig.~\ref{kiccomp} the temperatures derived from spectroscopy show reasonable agreement with the ones quoted in the KIC, agreeing within the KIC uncertainty of $\pm200$ K. The KIC \teff\ appears to be systematically lower when moving towards the hotter stars in our sample, as indicated by the linear fit. A similar trend is seen for the \feh\ values, with the KIC values being systematically lower for lower \teff. The discrepancies of the \logg\ values are more evenly scattered around a negligible offset of 0.003 dex. The scatter of both \feh\ and \logg\ is large as seen in the figure, but taking the KIC uncertainty of $\pm0.5$ dex on \logg\ into account the agreement is reasonable, except a few outliers. There are no uncertainty quoted in the KIC on the \feh\ for giants \citep{brown2}, but it is expected to be large. Furthermore we find very large discrepancies for individual stars of more than 2.0 dex in both \logg\ and \feh, which stresses the need for detailed spectroscopic follow-up of \kepler\ targets. These extreme outliers in \logg\ can likely be explained by a misclassification of the stars in the photometric analysis done for the KIC. As discussed by \citet{brown2}, subgiants can be misclassified as dwarf stars in the KIC, thus leading to wrong estimates for the surface gravity. We note that 10 of the stars in our sample do not have any values in the KIC. Below we give the coefficients on the linear fits shown in Fig.~\ref{kiccomp}. We note that the offset and RMS-scatter given in the figure is calculated with respect to the linear fits:

\begin{figure}[htb]%
\includegraphics[width=\columnwidth]{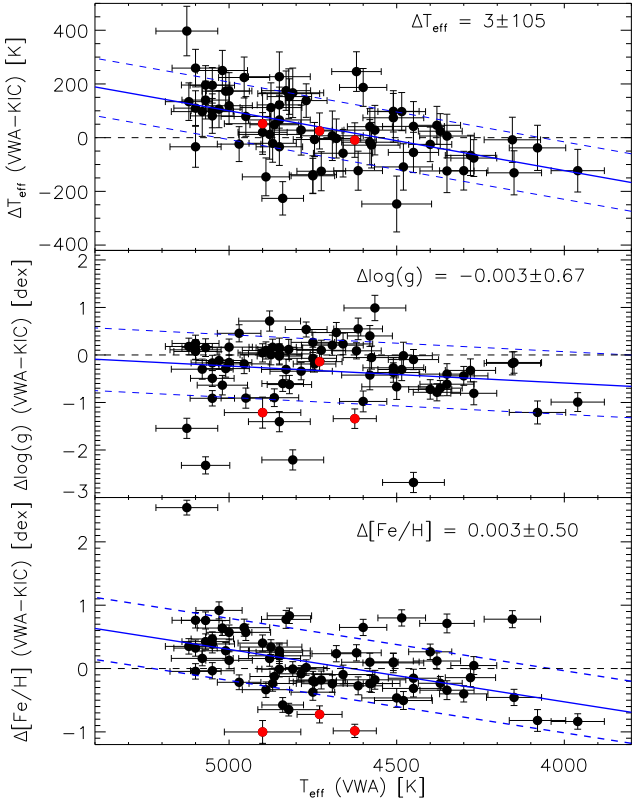}%
\caption{Comparison between parameters determined by VWA and the values found in the KIC. The Pop. II stars in our sample are marked with red circles. The devations and RMS-scatter are calculated with respect to the linear fits shown as solid lines. The dashed blue lines indicate the RMS-scatter.}%
\label{kiccomp}%
\end{figure}

\begin{eqnarray}
\Delta T_{\rm{eff}} & = & 0.22T_{\rm{eff}} - 1013\\
\Delta\log(g) & = & 3.60\times 10^{-4} T_{\rm{eff}} - 2.02\\
\Delta \rm{[Fe/H]} & = & 8.27\times 10^{-4} T_{\rm{eff}} - 3.83.
\end{eqnarray}

\subsection{Utilizing the asteroseismic \logg}
\label{utillog}
Determining \logg\ from spectroscopy is difficult because a change in \teff\ affects how the derived \feone-abundances depend on lines with different EP, and changes in \logg\ and \teff\ strongly affect the ionization equilibrium of \feone\ and \fetwo. This can result in a strong correlation between \logg\ and \teff\ in the spectroscopic analysis. The asteroseismic \logg\ should be a more precise and robust measure of the surface gravity as the asteroseismic \logg\ only depends on the gross properties of the star \citep{gai}. We use the method of \citet{kallinger} to determine the fundamental parameters for 60 stars from the \kepler\ observations. For an additional 21 stars we found fundamental parameters using the scaling relations of \citet{white}. We were unable to derive asteroseismic parameters of one giant, KIC 9574235, as it oscillates at such a low frequency that an analysis requires a longer timeseries than is currently available.

\begin{figure}[htb]%
\includegraphics[width=\columnwidth]{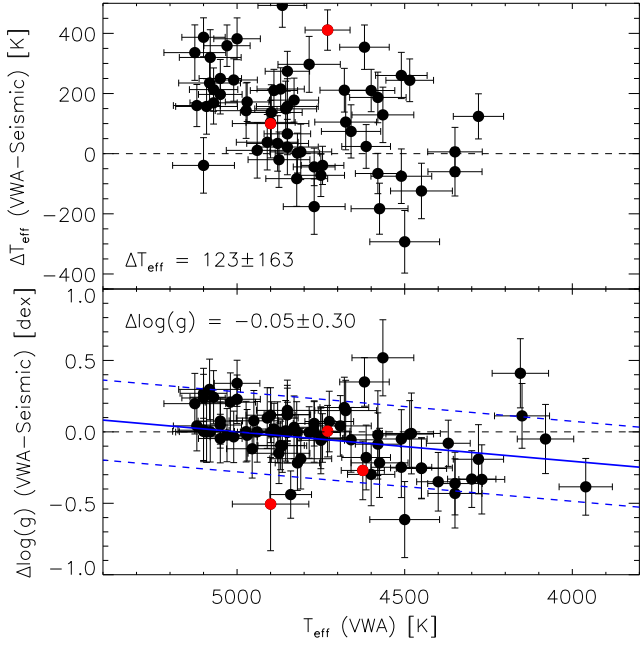}%
\caption{Comparison between parameters determined by VWA and from the asteroseismic analysis. The Pop. II stars in our sample are marked with red circles. The mean deviations and RMS-scatter are shown in the figure. The blue line indicate a linear fit to the deviation in \logg\ with the dashed lines indicating the RMS-scatter of the fit. For clarity, one strong outlier in $\Delta$\logg\ is not shown in the plot.}%
\label{seiscomp}%
\end{figure}

In Fig.~\ref{seiscomp} we compare for 81 giants the spectroscopic parameters to the parameters derived from the asteroseismic analysis. In the top panel we compare the \teff's determined from the two methods. A large scatter of the points is observed and we find a mean deviation of $123\pm163$ K. The scatter of the points is too large to reliably determine if any trends are present, but the \teff's derived from the model grid used by \citet{kallinger} tends to obtain lower values compared to the spectroscopic results. In the bottom panel of Fig.~\ref{seiscomp} we compare the \logg\ determination from the two methods. Good agreement is observed between the asteroseismic and spectroscopic \logg, with the spectroscopic and asteroseismic \logg\ agreeing within $1\sigma$ in more than half the cases. The mean deviation is found to be $-0.05\pm0.30$ dex. 

\begin{figure}[htb]%
\includegraphics[width=\columnwidth]{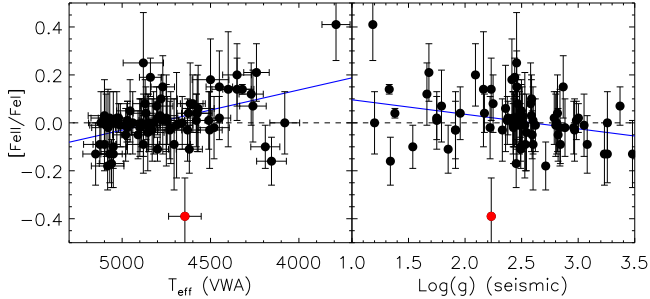}%
\caption{[\fetwo /\feone] vs. \teff\ (left) and \logg\ (right) for our analysis with \logg\ fixed at the asteroseismic value. The discrepancy correlates with both \teff\ and \logg, increasing towards lower \logg. One outlier is shown in red.}%
\label{fe2fe1}%
\end{figure}

We see discrepancies in \logg\ that correlates with the temperature which likely originates from using ionization equilibrium of \feone\ and \fetwo\ to determine the spectroscopic \logg. At the temperature and gravity ranges considered here, one would expect non-local thermodynamic equilibrium (NLTE) effects to influence the \feone\ abundances. Neglecting NLTE effects will lead to an underestimation of the abundance of \feone, because the model \feone\ lines become stronger when NLTE is not taken into account. This will in turn be reflected in the discrepancy between \logg\ derived from asteroseismology and spectroscopy. An illustration of this is seen in Fig.~\ref{fe2fe1} where we have fixed \logg\ at the asteroseismic values in our spectroscopic analysis. This produces a disagreement between the \feone\ and \fetwo\ abundance that correlates with the temperature as well as with \logg. This effect increases with decreasing \logg. The linear fits in Fig.~\ref{fe2fe1} merely serve as indication of systematic trends in the abundance differences and should not be used to calibrate NLTE effects. As shown by \citet{mashonkina} the NLTE corrections can vary by as much as 0.16 dex between different iron lines in the same model, so this has to be treated on a line-by-line basis. Although the NLTE corrections from \citet{mashonkina} are for a different parameter space than in this work, the effects of NLTE are expected to increase for the giants.

\begin{figure}[htb]%
\includegraphics[width=\columnwidth]{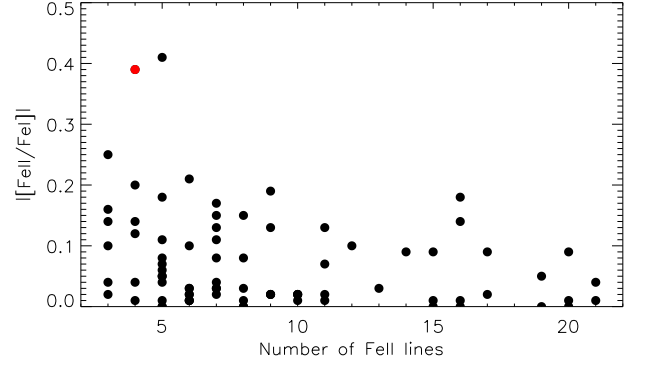}%
\caption{Absolute value of the abundance difference [Fe\,{\sc I}/H]-[Fe\,{\sc II}/H]=[\fetwo/\feone] vs. number of \fetwo\ lines used in the analysis. Removing one outlier (red), there is no obvious correlation between the difference and the number of lines used in the analysis.}%
\label{fevsnofe}%
\end{figure}

Another explanation for this disagreement could be that some of the giants have as few as three \fetwo\ lines, which could produce a less robust determination of \logg\ from the ionization equilibrium. However, inspecting Fig.~\ref{fevsnofe} after removing one outlier, which was also an outlier in Fig.~\ref{fe2fe1}, there is no compelling evidence that the number of lines has any significant influence on the discrepancy between the two iron species. For these reasons we adopt the asteroseismic \logg\ value when determining fundamental parameters. No NLTE departures are expected for \fetwo\ at these values of \logg\ and \teff, so we adopt the \fetwo\ abundance as the best measure for \feh\ (L. Mashonkina, 2011, priv. communication). 

With the \logg\ fixed at the asteroseismic value we adjusted the \teff\ and microturbulence to remove EP and EW correlations. The values of \teff\ and microturbulence determined this way are used for the remainder of this paper.

One extreme outlier between the \logg\ derived from spectroscopy and asteroseismology is the giant KIC 4070746. This star shows up as an outlier both in comparison with the KIC as well as with asteroseismic results. While a comparison of the spectrum of KIC 4070746 with the spectrum of a similar star in terms of \teff\ and \feh\ suggests that the spectroscopic value of \logg\ is reliable, the power spectrum shows clear power excess around 190 $\mu$Hz, consistent with the \logg-value determined from asteroseismology. We note that this target has the lowest S/N spectrum of all stars in our sample ($\leq70$) and very few useful iron lines. This, combined with the difficulty of proper continuum determination in the observed spectrum, can affect the spectroscopic parameter determination. We thus adopt the asteroseismic \logg-value for this star as well.

\subsection{Photometric calibration of \teff}
\begin{figure}%
\includegraphics[width=\columnwidth]{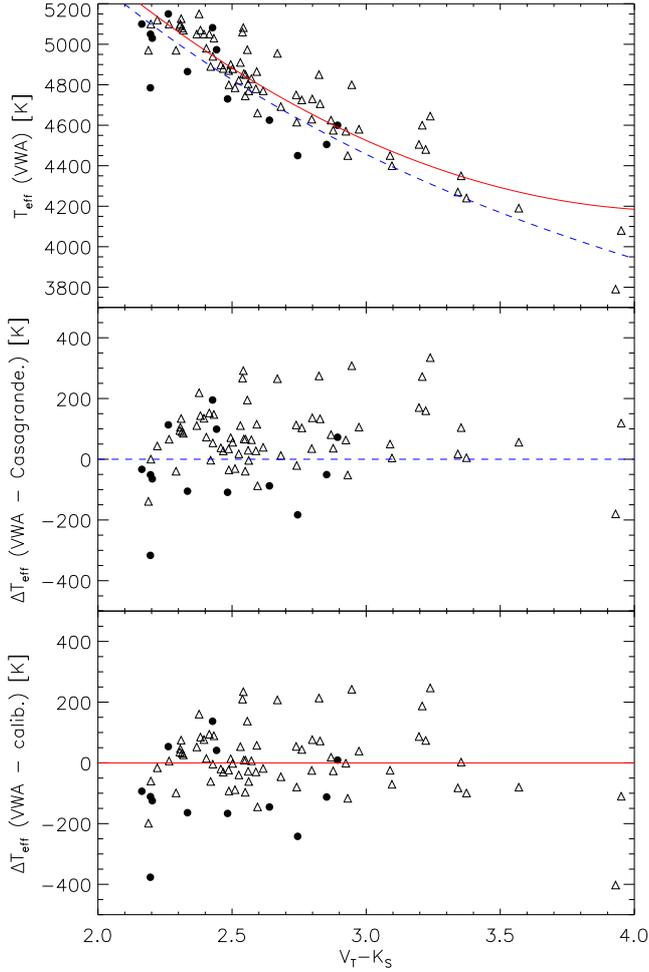}%
\caption{The upper panel shows the spectroscopic \teff\ vs. $(V_T-K_S)$ for the giants in our sample. The filled circles show targets for which we were able to measure the reddening, whereas the open triangles have not been corrected for reddening. The solid red line shows the calibration from in \citet{bruntt5} based on 93 main sequence and sub-giant stars. The dashed blue line shows the calibration from \citet{casagrande}. The two lower panels show the residuals.}%
\label{photcalib}%
\end{figure}

We used the spectroscopic \teff\ to test the calibration of the temperature scale by \citet{bruntt5}, which is valid for main-sequence and sub-giant stars. The calibration is based on the photometric index $(V_T-K_S)$, where $V_T$ is the TYCHO $V$-magnitude and $K_S$ is the $K$-magnitude from the 2MASS photometry. This index has previously been shown by \citet{casagrande} to be a very good indicator of \teff. Fig.~\ref{photcalib} shows that the Bruntt calibration fits the giants well. However, interstellar reddening potentially influences the color index of the stars and needs to be taken into account. \citet{bruntt5} found that interstellar reddening was negligible for the main-sequence stars investigated in their work. The same was also reported from standard photometric observations by \citet{molenda2}. However, reddening is expected to play a more significant role for the giants. 

In order to investigate this, we measured EW's for the interstellar Na D1 line in 19 giants where this line could be separated from the stellar Na line. The reddening was estimated using the calibration of \citet{munari}. Significant reddening is found for most of the giants where this analysis could be performed, reaching as high as $E(B-V)=0.27$. The reddening was transformed to the $(V_T-K_S)$ index using the transformation of \citet{ramirez}. 

Of the 19 targets where reddening could be measured, six targets were excluded from the photometric calibration because they were too faint ($V_T\geq12$) making the photometry unreliable \citep{hoeg}. In Fig.~\ref{photcalib}, open symbols are used for targets without reddening corrections and the filled circles for the 13 targets where we measured the reddening. The general agreement between the spectroscopic temperatures and the calibration of \citet{bruntt5} as well as the calibration of \citet{casagrande} appears to be good, but might worsen if reddening corrections are taken into account, which will potentially shift the photometric index by a non-negligible amount. Without knowledge about the reddening for a larger sample of the giants we cannot draw a clear conclusion about the validity of either temperature calibration. We plan to collect multi-color photometry to address the reddening problem for the remaining stars in our sample.

\subsection{Calibration of microturbulence}
\citet{bruntt5} provided a calibration of the microturbulence based on \logg\ and \teff\ for a large sample of main-sequence and subgiant stars. We have tested the calibration on our sample of stars, but find that for the giants the fitting coefficients need to be adjusted. We found that the following calibration gave the lowest residual:

\begin{eqnarray}
\xi_t/ \rm{km\ s}^{-1} & = &0.871 - 2.42\times10^{-4}(T_{\rm{eff}}-5700) \nonumber \\
& &- 2.77\times10^{-7}(T_{\rm{eff}}-5700)^2 \nonumber \\
& &- 0.356\times(\log g - 4.0).
\label{vmiccalib2}
\end{eqnarray}

\begin{figure}%
\includegraphics[width=\columnwidth]{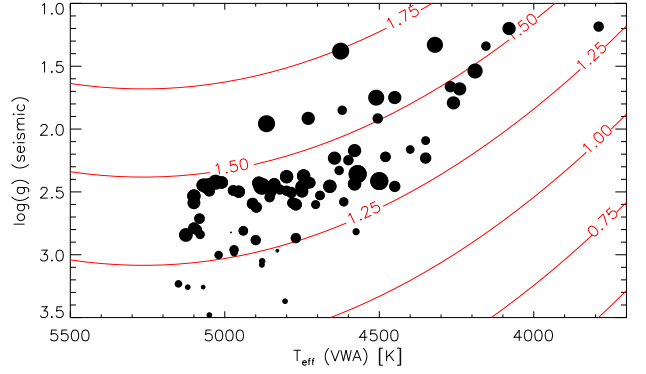}%
\caption{Diagram of 81 giants with asteroseismic \logg\ in the (\logg, \teff)-plane. Overplotted are contours of constant microturbulence (Eq. \ref{vmiccalib2}). The size of the symbols is proportional to the microturbulence as determined from the spectroscopic analysis. The microturbulence is in the range [0.9,2.25] \kms.}%
\label{vmic}%
\end{figure}

This calibration gives an RMS-scatter of the residuals of 0.12 \kms\ which is 30\% lower than with the calibration presented in \citet{bruntt5}. Taking into account a typical uncertainty of 80 K on \teff\ and 0.20 dex on \logg, the total uncertainty on the microturbulence from this calibration is found to be 0.14 \kms. In Fig.~\ref{vmic} we plot the stars in our sample in the (\logg, \teff)-plane, overplotted with contours of constant microturbulence, using the calibration of Eq.~\ref{vmiccalib2}.

\section{The abundance pattern}
After the fundamental parameters had been determined using the iron lines, we derived mean abundances for several other elements. The abundances for all targets are presented in Table~\ref{abundseis}. We use the models with the \logg\ fixed at the asteroseismic values. For one star where no asteroseismic parameters were available, we used the parameters provided by the standard spectroscopic analysis with \logg\ determined from ionization equilibrium. All abundances are measured relative to the Sun, using the \citet{grevesse} values for the solar abundances, because their work is used as the reference abundances in the MARCS models as well as for the calculation of the log-$gf$ corrections used in our spectroscopic analysis.

\begin{figure}%
\includegraphics[width=\columnwidth]{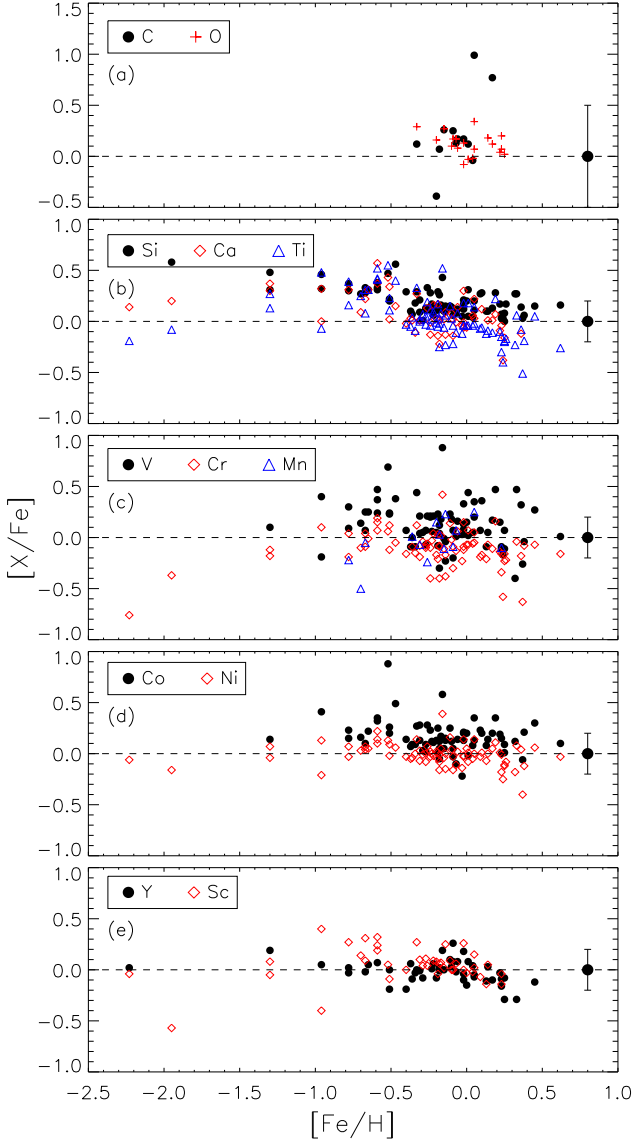}%
\caption{[X/Fe] vs. \feh\ for all heavy elements measured in the giants for which we were able to determine an asteroseismic \logg. \fetwo\ has been used as a proxy for the iron abundance. The errorbars located at \feh\ $=0.8$ are representative for the uncertainties on the abundances in each panel.}%
\label{abundfig}%
\end{figure}

Carbon and oxygen have been detected in the stars where NARVAL/ESPaDOnS spectra were available, due to their higher S/N as well as their more extended wavelength coverage. These targets all display roughly solar metallicity, so no correlations with metallicity can be determined from this dataset. As seen in Fig.~\ref{abundfig}(a), a weak over-abundance of both elements can be observed, compared to iron. We note that the abundances are associated with large uncertainties of roughly 0.5 dex, as only 3-4 atomic lines were available for the abundance determination. Within these uncertainties the abundances scale with the iron abundance in the giants. The two very carbon-rich giants are not significantly different from the rest of the giants with carbon detected, in terms of \teff, \logg\ and \feh. Even though they display the largest individual uncertainties ($\pm0.8$ dex), inspection of the spectra shows strong absorption in the C-lines of these giants, compared to the giants with solar carbon abundance. 

In Fig.~\ref{abundfig}(b) we present results for the $\alpha$-element abundances. Si appears overabundant for most targets whereas Ca and Ti scales roughly with the iron abundance for \feh\ $\geq-0.4$. Towards lower metallicities the abundances increase for decreasing metallicity. This is consistent with what is observed by \citet{bruntt5} for the 93 main-sequence stars in their work.

In Fig.~\ref{abundfig}(c-d) we show the abundances for the iron peak elements, V, Cr, Mn, Co and Ni. Ni scales with the iron abundance whereas V and Co appear overabundant, compared to Fe. The overabundance of Co can be explained by the presence of hyperfine splitting of the spectral lines. This is not taken into account in VWA and leads to an over-estimation of the abundance. A study of cobalt abundances in 29 red giants by \citet{boyarchuk} shows that neglecting this effect results in an average overestimation of the abundance of $0.08$ dex. If this is substracted from the abundances derived by VWA, Co scales with the Fe abundance, within the uncertainties. However, the subtraction has not been done for the abundances quoted in Table~\ref{abundseis} as large star-to-star differences from hyperfine splitting are observed in the work of \citet{boyarchuk}.

For V the picture is less clear. A study by \citet{puzeras} of 62 red clump stars also shows vanadium to be overabundant by $0.11\pm0.12$ dex whereas a study by \citet{luck} find $-0.07\pm0.20$ dex for their sample of giants. From this, there appear to be no clear trend for giants, but within the uncertainties, both studies find V to scale with the Fe abundance. This is also the case for the present analysis.

Cr appears underabundant, relative to Fe, which is a well-known feature for giants (see for instance \citealt{johnson}). Standard spectroscopic analysis \citep{johnson} also finds a systematic difference between Cr\,{\sc I} and Cr\,{\sc II} abundances, but this is only observed for three of the 21 giants in our work, where both species are detected. The Mn abundance is only measured for a handful of stars, so we cannot make any clear conclusion about general trends.

Finally in Fig.~\ref{abundfig}(e), are shown the abundances of Y and Sc, which are associated with the  slow and rapid neutron-capture process respectively. Both elements appear to scale roughly with the iron abundance. 

\section{Influence of metallicity on asteroseismic properties}
\label{seismo}
In order to facilitate the modeling of the giant stars in this paper, we have tried
to select pairs of stars ({\it twins}) with similar parameters (\teff, $\Delta\nu$, $\nu_\mathrm{max}$) except for metallicity.
The criteria we have used for selecting these {\it twins} was a rather
naive idea about how similar stars could be found.  

{\it Twins} were identified by calculating the deviation, $\sigma_\mathrm{tot}$, between the large frequency separation, $\Delta\nu$, frequency of maximum power, $\nu_\mathrm{max}$ and the effective temperature, \teff\ for a target star (tgt) and a potential {\it twin} (tw) in the following way:
\small
\begin{equation}
\sigma_\mathrm{tot}= \sqrt{\left(\frac{\Delta\nu_\mathrm{tgt} - \Delta\nu_\mathrm{tw}}{\sigma(\Delta\nu_\mathrm{tgt})}\right)^2 + \left(\frac{\nu_\mathrm{max,tgt} - \nu_\mathrm{max,tw}}{\sigma(\nu_\mathrm{max,tgt})}\right)^2 + \left(\frac{\teffii,_\mathrm{tgt} - \teffii,_\mathrm{tw}}{\sigma(\teffii,_\mathrm{tgt})}\right)^2}
\label{twinsel}
\end{equation}
\normalsize
Here a small $\sigma_\mathrm{tot}$ would indicate a close {\it twin} in terms of \teff\ and oscillation parameters, 
but since we were interested in {\it twins} with different metallicities, we defined an additional 'quality factor' 
for the {\it twins} in the sample as

\begin{equation}
\centering
Q_\mathrm{tw} = \frac{\Delta\mathrm{[Fe/H]}}{\sigma_\mathrm{tot}}
\end{equation}

\noindent
where $\Delta\mathrm{[Fe/H]}$ was the difference in KIC metallicity between a target star and a potential {\it twin}. Hence a large $Q_\mathrm{tw}$ was taken as indicating a good target for our investigation. However, this approach does not
take into account that there is a correlation between \feh\ and \teff\ which limited the selection of {\it twins} with very different metallicities.

To give an idea what the power spectra of sets of {\it twins} look like,
we present two sets of {\it twins}. The first set consists
of two RGB stars (H burning). The classification is based on the average period spacing
$\Delta P$, of the dipole modes, measured from the power spectra as described by \citet{bedding1}. The second set presents two
red clump (RC) stars (He burning) again classified by the measured average 
value of $\Delta P$.
The parameters for the stars are summarized in Table~\ref{parameters}.
The asteroseismic parameters $\Delta\nu$ and $\nu_\mathrm{max}$ and their uncertainties are derived as the mean values and the scatter on the results from a number of different pipelines (see \citet{hekker3} for a discussion). The \teff\ and \feh\ are from the spectroscopic analysis in this paper and the mass, $M$, is obtained from the scaling laws of \citet{white}.

\begin{table*}
\caption[ ]{Parameters for two sets of giant {\it twins}. The first set is RGB stars and the last set red clump stars.} \label{parameters}
\begin{center}
\begin{tabular}{lllllll}  \hline
KIC \# & $\Delta\nu$ $[\mu]$Hz & $\nu_\mathrm{max}$ $[\mu]$Hz & $\Delta P$ [s] & \teff\ [K] & $M/$\msun & \feh \\
\hline
3744043 & $9.86\pm0.24$ & $110.25\pm6.09$ & 58 & 4970 & 1.21 & -0.31 \\
6690139 & $9.65\pm0.22$ & $115.33\pm6.28$ & 49 & 5020 & 1.55 & -0.13 \\
\hline
11444313 & $3.97\pm0.14$ & $33.62\pm2.56$ & 272 & 4750 & 1.16 & -0.01 \\
11569659 & $4.09\pm0.13$ & $31.85\pm2.58$ & 245 & 4890 & 0.95 & -0.27 \\
\hline
\end{tabular}
\end{center}
\end{table*}

\begin{figure*}
\centering
\includegraphics[width=15cm]{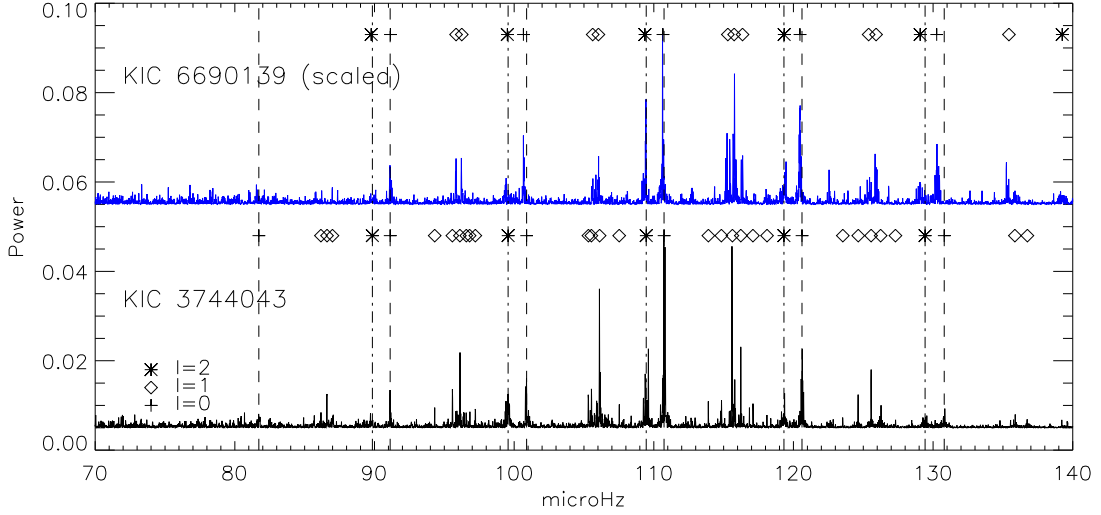}%
\caption{The power spectra for a set of RGB {\it twins}. The main difference between the stars is the
metallicity. The position of the $\ell = 0,2$ modes for the bottom spectrum is 
indicated by vertical lines (dash and dot-dash respectively). Modes fitted with the MCMC code (see text) are marked
by symbols at the top of each spectrum and the meaning of the symbols are
displayed at the bottom left corner. The power spectrum of the star in the top panel has been scaled using the ratio of the $\Delta\nu$'s of the two stars.}
\label{rgbfig}%
\end{figure*}

\begin{figure*}
\centering
\includegraphics[width=15cm]{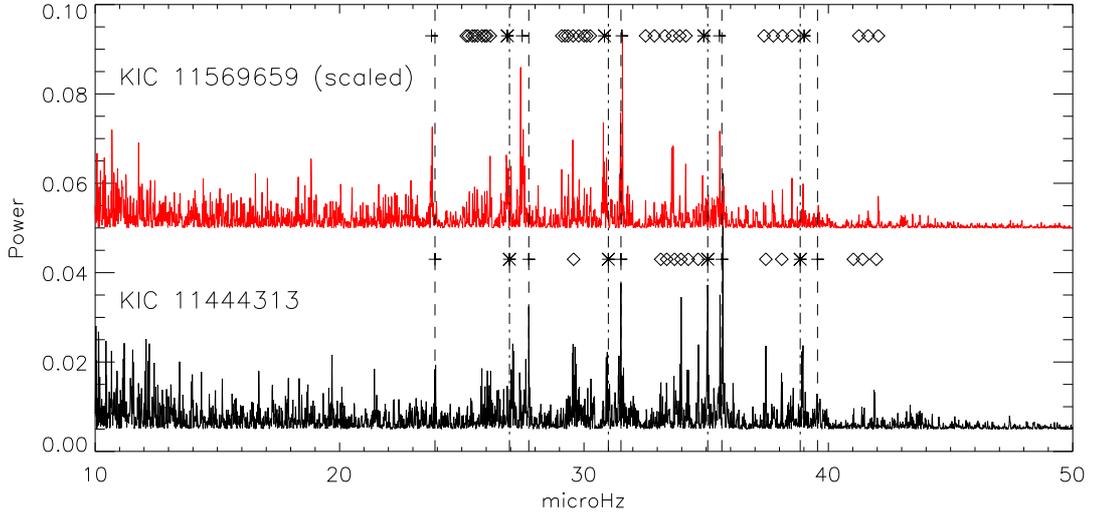}%
\caption{Same plot as Fig.~\ref{rgbfig} for a set of red clump {\it twins}.}
\label{rcfig}%
\end{figure*}

It is not the intention here to do a detailed modeling of the two sets of
{\it twins}. Instead two figures present the power spectra of the \kepler\ 
timeseries including data up to observing quarter 7 (Q7) corresponding to 494 days of observing. Even though the {\it twins} have similar
parameters, the power spectra are noticably different. One star is shown without any
frequency scaling. The power spectra for the other star have been rescaled
(see \citealt{bedding2})
using a factor close to the ratio between the large frequency separations
$\Delta\nu$. We use the ratio from the $\Delta\nu$'s in Table~\ref{parameters} as a first guess, but in order to match
the central $\ell = 0$  modes more accurately, the ratio
has to be finetuned with a precision better than 1\%. The scalings applied were
0.978 and 1.024 in Figs.~\ref{rgbfig} and \ref{rcfig} respectively.
The scaling factor determines the mean density ratio of the stars (\citealt{bedding2}).

To obtain estimates of the mode degree and frequencies we applied
the Markov Chain Monte Carlo (MCMC) technique by \citet{handberg}. The modes are marked
in Figs.~\ref{rgbfig} and \ref{rcfig} by different symbols.

The two targets in Fig.~\ref{rgbfig} have similar power spectra. The scaling of the power spectrum aligns the $\ell=0,2$ modes well. Also, as is evident, there are less mixed modes in the top panel, where the mixed modes seem to be located within a narrower frequency range.  
A detailed analysis is needed to determine if the differences are
due to different masses or metallicities. 

The same similarities and discrepancies are present for the second set of {\it twins} (see Fig.~\ref{rcfig}). The power spectra look very
similar, but even the $\ell = 0$ modes can not be made to match by
a simple scaling. Again the $\ell = 1$ modes differ and constitute
such a rich set that the identification of modes becomes really 
difficult as the $\ell = 1$ modes start to overlap with modes of other degrees.

Interestingly, the simultaneous analysis of scaled power spectra 
makes it easier to identify
more modes as they line up nicely and modes that one might discard in a single
spectrum now appear to be 'significant' when they match up with a mode in a
similar spectrum of a similar star as first pointed out by \citet{bedding2}.

\begin{figure*}[htb]
\centering
\includegraphics[width=\textwidth]{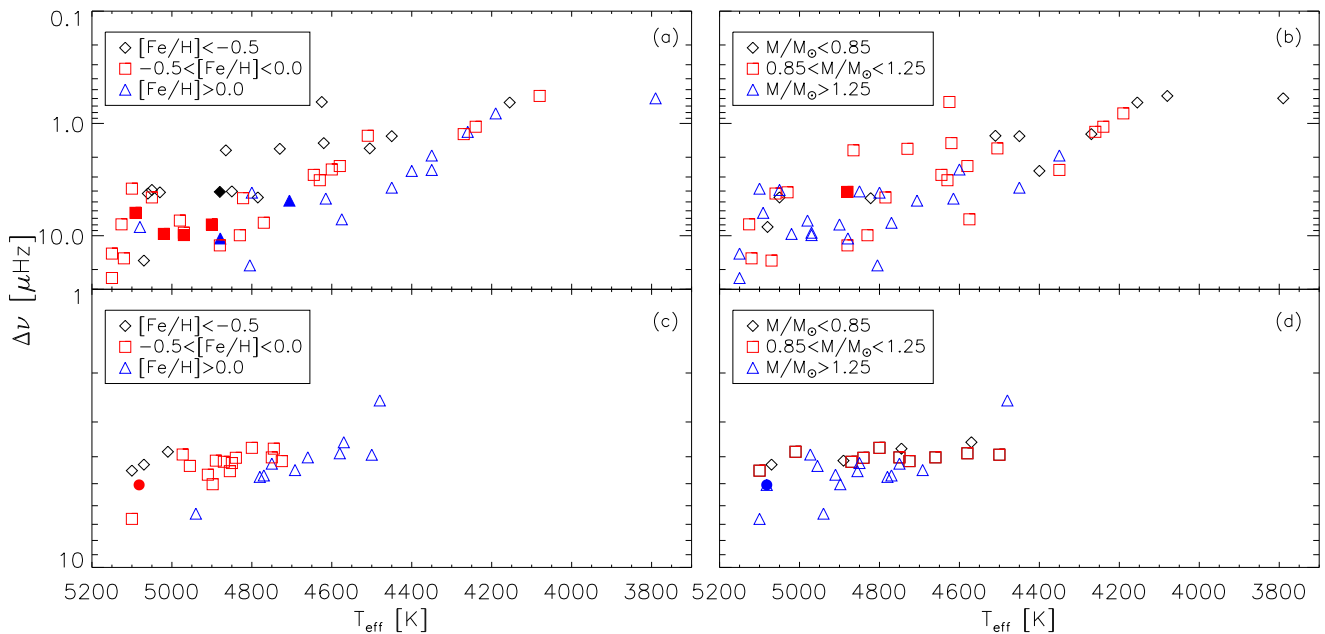}%
\caption{Panels (a) and (b) show $\Delta\nu$ vs. \teff\ for the giants that could not be identified as clump stars based on $\Delta P$. In panel (a) the stars have been color-coded according to metallicity. Black diamonds are \feh\ $<-0.5$, red squares are $-0.5\leq$ \feh\ $<0.0$ and blue triangles $0.0<$ \feh. In panel (b) the stars have been color-coded according to mass. Filled symbols are stars that are confirmed RGB stars based on $\Delta P$. Panels (c) and (d): Same as top panels, but for the red clump stars, identified from their $\Delta P$-value. The filled symbols show the one star identified as belonging to the secondary clump. Note the different scalings on the axes in the top and bottom panels.}%
\label{dnuclump}%
\end{figure*}

A clear metallicity effect is visible when using the asteroseismic parameters in combination with spectroscopy (Fig.~\ref{dnuclump}). Here we have split the stars in two subsets, based on mass and their period spacing, $\Delta P$. Top panels show stars that belong to the RGB population. We have removed targets with $\Delta P > 100\, \mathrm{s}$ as they belong to the RC, following the procedure of \citet{bedding1}. Also, all stars with a mass larger than 2 \msun\ have been removed, as we assume they belong to the secondary clump (SC). The seven stars plotted with filled symbols in the top panels are confirmed RGB stars, based on their period spacing.  We assume that the remaining targets belong to the RGB as well, as the RC stars occupy a very narrow range in \teff\ and $\Delta\nu$, although we cannot strictly rule out that a few RC giants are present in the top panels. The stars in panels (a) and (c) have been color coded according to their metallicity. We use three bins with \feh\ $<-0.5$ (black diamonds), $-0.5\leq$ \feh\ $<0.0$ (red squares) and $0.0<$ \feh\ (blue triangles). In Fig.~\ref{dnuclump}(a) it can be seen that the RGB stars shift to lower values of \teff\ as the metallicity increases, in the $\Delta\nu$ vs. \teff\ diagram, splitting into three metallicity branches. 

In the bottom panels of Fig.~\ref{dnuclump} we present the stars in the sample that could be identified as RC giants, based on their values of $\Delta P$. In panel (c) it is clear that also the RC stars separate into three groups. It appears more pronounced because the range in $\Delta\nu$ is much smaller in the $\Delta\nu$-\teff\ plane.

\begin{figure}[htb]
\centering
\includegraphics[width=0.85\columnwidth]{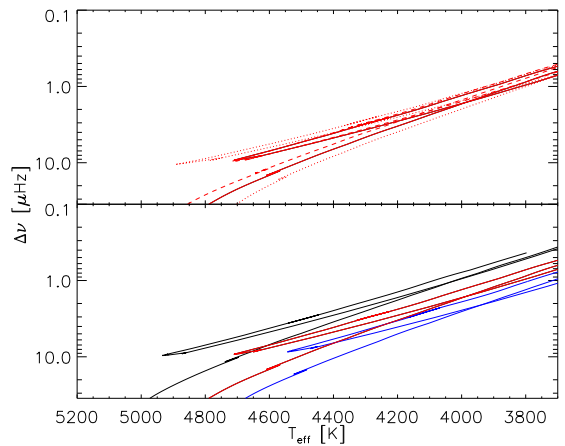}%
\caption{Top panel: Evolutionary tracks, showing the change in $\Delta\nu$ for a model of solar metallicity and mass equal 0.8 \msun\ (dotted), 1.0 \msun\ (solid) and 1.2\msun\ (dashed). Lower panel: Evolutionary tracks of a 1 \msun\ model with \feh\ of $-0.35$ dex (black), 0.06 dex (red) and 0.4 dex (blue).}%
\label{evoltrack}%
\end{figure}

By looking at evolutionary tracks we can compare the observations in Fig.~\ref{dnuclump} to stellar models. In Fig.~\ref{evoltrack} we plot evolutionary tracks of different metallicity and mass. In the top panel we show models with solar metallicity and masses 0.8, 1.0 and 1.2 \msun, consistent with the mass-range of the giants investigated here. In the lower panel we show evolutionary tracks for 1 \msun\ with \feh\ $=-0.35$, 0.06 and 0.4 dex. The evolutionary tracks are taken from the BaSTI database \citep{cassisi}, from which we calculate $\Delta\nu$, using the scaling relations of \citet{white}. From the models, the influence of metallicity differences is expected to be larger than mass differences for the RGB stars. 

For RC giants, the effects of metallicity differences and mass differences should be comparable for the low-mass stars according to the evolutionary tracks, but dominated by metallicity differences for masses higher than 1 \msun. Thus, judging from Fig.~\ref{dnuclump}(c) the observed separation may as well be related to mass differences for stars with $M<1.0$ \msun.

In Fig.~\ref{dnuclump} panels (b) and (d) is shown the RGB and RC giants color coded according to mass rather than metallicity. It is clear that the two populations do not show any systematic behaviour with mass. Thus, we conclude that the systematic behaviour of both RGB and RC giants in Fig.~\ref{dnuclump} (a) and (c) is a metallicity effect rather than an effect of different masses. As discussed above, this is consistent with the evolutionary tracks for RGB stars, but in contrast to what is expected for the low-mass RC stars. Detailed modeling will be needed to determine the exact cause of this, which is beyond the scope of this paper. 

This separation in metallicity could potentially open up for a rough estimation of the metallicity of RC stars based on a combination of asteroseismic and spectroscopic parameters without a detailed abundance study.

A number of both RGB and RC stars are seen to have very low masses ($<0.85$ \msun), which may likely be underestimated. As shown by \citet{miglio2} uncertainties on red giant masses derived from the scaling relations alone, have uncertainties of up to 15\%. Taking this into account does, however, not change the conclusion that the observed separation in Fig.~\ref{dnuclump} is dominated by metallicity. We note that such low masses are also observed in a fraction of the more than 10,000 public \kepler\ red giants (Stello et al., in preparation).

\section{Conclusions}
In this paper we have presented fundamental parameters as well as detailed abundance analysis of 82 red giants that are targets of the NASA \kepler\ mission, and we have identified some effects of metallicity on the stellar oscillations. The \teff\ found in the KIC is in reasonable agreement with the spectroscopic one, but the KIC \teff\ appears to be systematically lower than the spectroscopic value, when moving towards hotter stars. The values for \logg\ and \feh\ from the KIC are more unreliable for the giants, than is the case for main-sequence stars, with an RMS-scatter of 0.67 dex for \logg\ and 0.50 dex for \feh, and we identify serious individual discrepancies in both parameters, in the worst cases higher than 2.0 dex. We test a photometric calibration of \teff\ from the photometric index ($V_T-K_S$), presented in \citet{bruntt5}, but information about the reddening of the giants is needed in order to address the validity of this calibration. Furthermore, we test a calibration of the microturbulence for dwarf and sub-giant stars from the same paper, but find that a different calibration gives a better fit to the observed values of the giants.

The comparison of spectroscopic and asteroseismic parameters indicates that the \teff\ from the model grid used in the astroseismic analysis is underestimated compared to the spectroscopic \teff\ (see Fig.~\ref{seiscomp}). However, this difference can likely be explained by differences between the model grids used for the asteroseismic and spectroscopic analyses.

For more than half the stars in our sample, agreement was found between the asteroseismic and the spectroscopic \logg. The mean offset of $-0.05$ dex is negligible compared to the RMS-scatter of 0.30 dex. For lower values of \logg\ disagreement is found between the two methods. We attribute this to NLTE effects acting in the atmospheres of the giants with low values of \logg\ ($\leq2.5$ dex), resulting in an erroneous \logg\ from spectroscopy. We have no means of addressing the magnitude of these effects, but note that an increase of NLTE effects is expected when moving to lower surface gravities, which is consistent with the general trend we observe.

The abundances of most elements appear to scale roughly with the iron abundance and we find that \feh\ describes the metallicity of most of the targets analysed here. However, we do observe a trend of increasing $\alpha$-element abundances when moving to lower values of \feh, as expected for metal-poor stars. For detailed modeling of low-metallicity giants, we advice that individual abundance patterns are measured from spectroscopy and used in the modeling. We note that the abundances derived from spectroscopy only measure the abundances present in the outer layers of the stars and that we have no means of addressing the He and H abundance in the giants from the available spectroscopy.

We identify five very metal-poor giants (\feh\ $<-1.2$ dex) in our sample, three of which we classify as Population II stars. These should present interesting targets for a detailed asteroseismic analysis. 

We provide parameters for 10 giants for which no values exist in the KIC. This is important for the the asteroseismic analysis, as few constraints can be put on the asteroseismic models if the fundamental atmospheric parameters are unknown. 

The results presented here will facilitate the asteroseismic investigations of a modest sample of bright giants in the \kepler\ field covering a range of parameters. It shows the need for spectroscopic follow-up of the giants that are targets for the \kepler\ mission. However, our sample is just a fraction of the more than 10,000 giants for which oscillations have been detected (\citealt{hekker1}, Stello et al., in prep.). These large numbers of stars clearly require an automated spectroscopic analysis method. Our results, based on detailed spectroscopic analyses, would serve as an excellent benchmark for the developement of such fully automatic method.

We have identified a potential metallicity effect influencing the excited oscillation modes, particularly visible for the RC giants. We observe a separation with metallicity for both RGB and RC giants in the $\Delta\nu$ vs. \teff\ plane. These results call for a more detailed investigation, which will require both additional \kepler\ observations of these stars as well as detailed model calculations to understand the oscillation behaviour and how that is influenced by different masses and metallicities. It is clear that there is a lot to be learned from combining ground-based observations with the asteroseismic parameters that \kepler\ has provided.

\begin{acknowledgements}
AOT acknowledges support from Sonderforschungsbereich SFB 881 "The Milky Way System" (subproject A5) of the German Research Foundation (DFG). SH acknowledges financial support from the Netherlands Organisation of Scientific Research (NWO). TK is supported by the FWO-Flanders under project O6260 - G.0728.11. KB acknowledges support from the Carlsberg Foundation. DS acknowledges support from the Australian Research Council. The authors would like to thank the entire \kepler-team for their continued effort to ensure the success of this mission. This research took advantage of the SIMBAD and VIZIER databases at the CDS, Strasbourg (France), and NASA's Astrophysics Data System Bibliographic Services. Funding for this Discovery mission is provided by NASA's Science Mission Directorate. 
\end{acknowledgements}

% ++++++++++++++++++++++++++++++++++
% The bibliography file:
% ++++++++++++++++++++++++++++++++++
\bibliographystyle{aa}
\bibliography{giantspaper2} %% Name of .bib file
% ++++++++++++++++++++++++++++++++++
\newpage
\appendix
\onecolumn
\begin{landscape}
\section{Tables of results from spectroscopic analysis}
%\onecolumn
\small
\begin{longtable}{r|cccr|cc|ccr|crcccrr}
\caption{\label{atmostable}Fundamental atmospheric parameters for all red giants in our sample. First, we present the parameters determined from the spectroscopic analysis with VWA. Uncertainties on \teff, \logg, $\xi_t$ and \feh\ are 80 K, 0.2 dex, 0.15 \kms\ and 0.15 dex, respectively. \feh\ is given as the mean of the \feone\ and \fetwo\ abundances. The uncertainties on the asteroseismic determination of \teff\ and \logg\ are 100 K and 0.01 dex. For completeness we also show the values found in the KIC. Last columns are the values determined from VWA by fixing the gravity at the asteroseismic \logg. Here, \fetwo\ is used as the measure of \feh. Also shown are \vsini, $v_{macro}$ and radial velocity, $v_{rad}$, with associated uncertainties of 1.0 \kms, 1.0 \kms\ and 0.15 \kms. In the last column is shown the type of the giant, being red giant branch (RGB), red clump (RC) or secondary clump (SC) based on their period spacing.}\\
\hline\hline
       &\multicolumn{4}{c|}{VWA}          & \multicolumn{2}{c|}{Asteroseismic}  &  &   KIC    &      &\multicolumn{3}{c}{VWA + asteroseismic}  & &      &     &   \\
KIC$-$ID & \teff & \logg & $\xi_t$  & \feh & \teff & \logg & \teff & \logg & \feh & \teff & \feh & $\xi_t$ & \vsini\ & $v_{macro}$ & $v_{rad}$ & Type\\
\hline
\endfirsthead
\caption{Continued.}\\
\hline\hline
       &\multicolumn{4}{c|}{VWA}          & \multicolumn{2}{c|}{Asteroseismic}  &  &   KIC    &      &\multicolumn{3}{c}{VWA + asteroseismic}  & &      &     &   \\
KIC$-$ID & \teff & \logg & $\xi_t$ & \feh & \teff & \logg & \teff & \logg & \feh & \teff & \feh & $\xi_t$ & \vsini\ & $v_{macro}$ &$v_{rad}$ & Type \\
\hline
\endhead
\hline
\endfoot
\hline\hline
\endlastfoot
 1726211 & 5010 & 2.39 & 1.46 & $-$0.68 & 4765 & 2.42 & 4837 & 2.68 & $-$0.96 & 5010 & $-$0.67 & 1.46 & 0.5 & 4.0 & $-$145.8 & RC \\
 2425631 & 4600 & 1.95 & 1.35 & $-$0.17 & 4390 & 2.25 & 4413 & 2.93 & $-$0.82 & 4600 & $-$0.04 & 1.35 & 1.8 & 1.9 &   19.1 & $-$ \\
 2714397 & 5000 & 2.68 & 1.50 & $-$0.40 & 4618 & 2.45 & 4881 & 2.52 & $-$0.53 & 5060 & $-$0.59 & 1.60 & 1.5 & 3.5 & $-$190.5 & $-$ \\
 3429205 & 5100 & 3.75 & 1.10 &  0.09 & $-$ & 3.48 & 4841 & 3.51 & $-$0.67 & 5050 & $-$0.11 & 1.10 & 2.1 & 2.0 &  $-$27.0 & $-$ \\
 3430868 & 5126 & 3.04 & 1.53 &  0.04 & 4790 & 2.84 & 4729 & 4.58 & $-$2.50 & 5126 & $-$0.06 & 1.53 & 5.1 & 0.6 &    5.4 & $-$ \\
 3644958 & 5000 & 3.30 & 1.30 & $-$0.17 & $-$ & 2.96 & 4826 & 3.46 & $-$0.74 & 4970 & $-$0.23 & 1.30 & 2.8 & 2.7 &    4.8 & $-$ \\
 3744043 & 4970 & 2.96 & 1.20 & $-$0.31 & 4798 & 2.98 & 4994 & 2.50 & $-$0.09 & 4970 & $-$0.31 & 1.20 & 2.5 & 1.5 &  $-$53.9  & RGB\\
 3748585 & 4615 & 2.40 & 1.30 &  0.10 & 4591 & 2.58 & 4738 & 1.85 &  0.37 & 4615 &  0.25 & 1.30 & 2.5 & 2.1 &   $-$5.8 & $-$ \\
 3748691 & 4750 & 2.45 & 1.45 &  0.10 & $-$ & 2.50 & 4892 & 2.18 &  0.47 & 4750 &  0.13 & 1.45 & 2.4 & 2.2 &   $-$0.1 & RGB\\
 3860139 & 4480 & 2.21 & 1.35 &  0.09 & $-$ & 2.22 & 4589 & 2.22 &  0.60 & 4480 &  0.10 & 1.35 & 4.0 & 2.5 &  $-$25.2 & RC \\
 3936921 & 4510 & 2.11 & 1.65 &  0.03 & 4585 & 2.36 & 4436 & 2.38 & $-$0.06 & 4570 &  0.09 & 1.75 & 2.0 & 5.0 &  $-$48.9 & RC\\
 3955590 & 4565 & 2.75 & 1.50 &  0.34 & 4436 & 2.23 & 4537 & 1.76 &  0.52 & 4645 & $-$0.16 & 1.50 & 2.1 & 2.1 &  $-$57.2 & $-$ \\
 4070746 & 5070 & 1.30 & 1.10 & $-$0.05 & 4857 & 3.23 & 4874 & 3.62 & $-$0.81 & 5150 & $-$0.17 & 1.20 & 2.2 & 2.2 &   $-$1.6 & $-$ \\
 4072740 & 4875 & 3.22 & 1.20 &  0.15 & 4895 & 3.37 & 4763 & 3.22 & $-$0.19 & 4805 &  0.23 & 1.10 & 3.0 & 2.2 &  $-$17.1 & $-$ \\
 4157282 & 4350 & 1.73 & 1.27 & $-$0.07 & 4344 & 2.09 & 4344 & 2.13 & $-$0.78 & 4350 &  0.22 & 1.27 & 2.0 & 3.0 &  $-$37.4 & $-$ \\
 4177025 & 4270 & 1.33 & 1.40 & $-$0.44 & $-$ & 1.66 & 4346 & 2.14 & $-$0.49 & 4270 & $-$0.24 & 1.40 & 4.5 & 2.0 & $-$123.1 & $-$ \\
 4262505 & 4900 & 3.00 & 1.35 & $-$0.16 & 4796 & 2.88 & 4880 & 2.96 & $-$0.56 & 4900 & $-$0.20 & 1.35 & 2.2 & 2.8 &  $-$14.0 & RGB\\
 4283484 & 5030 & 2.40 & 1.60 & $-$0.78 & 4671 & 2.42 & 4924 & 2.52 & $-$1.70 & 5030 & $-$0.77 & 1.60 & 3.2 & 2.7 &  $-$44.8 & $-$ \\
 4480358 & 4620 & 2.20 & 1.30 & $-$0.80 & 4266 & 1.85 & 4374 & 2.11 & $-$1.05 & 4620 & $-$0.96 & 1.30 & 2.1 & 1.8 &  $-$88.3 & $-$ \\
 4659706 & 4450 & 2.20 & 1.40 &  0.40 & $-$ & 2.46 & 4505 & 2.30 &  0.55 & 4450 &  0.62 & 1.40 & 3.1 & 2.9 &  $-$21.8 & $-$ \\
 5113061 & 4150 & 1.65 & 1.60 &  0.16 & $-$ & 1.54 & 4281 & 1.81 &  0.62 & 4190 &  0.01 & 1.60 & 2.3 & 2.3 &   $-$4.3 & $-$ \\
 5113910 & 4510 & 1.70 & 1.65 & $-$0.33 & 4250 & 1.75 & 4412 & 2.07 & $-$0.43 & 4510 & $-$0.31 & 1.65 & 3.1 & 3.1 &  $-$12.4 & $-$ \\
 5284127 & 4660 & 2.40 & 1.50 &  0.44 & 4586 & 2.46 & 4718 & 2.17 &  0.54 & 4660 &  0.45 & 1.50 & 3.5 & 2.8 &  $-$73.0 & RC \\
 5511423 & 4370 & 1.25 & 1.65 & $-$1.14 & $-$ & 1.33 & 4346 & 1.95 & $-$0.91 & 4320 & $-$0.96 & 1.65 & 1.8 & 1.9 &  $-$87.3 & $-$ \\
 5524720 & 4350 & 1.80 & 1.40 &  0.13 & 4410 & 2.23 & 4474 & 2.41 &  0.47 & 4350 &  0.38 & 1.40 & 3.0 & 3.1 &  $-$32.8 & $-$ \\
 5612549 & 4850 & 2.50 & 1.50 & $-$0.31 & 4784 & 2.38 & 4884 & 2.35 & $-$0.57 & 4800 & $-$0.33 & 1.50 & 3.8 & 3.4 &   $-$3.5 & RC\\
 5698156 & 4730 & 1.92 & 1.50 & $-$1.30 & 4319 & 1.92 & 4705 & 2.06 & $-$0.58 & 4730 & $-$1.33 & 1.50 & 3.9 & 3.2 & $-$381.2 & $-$ \\
 5700368 & 4850 & 2.50 & 1.33 & $-$0.15 & 4828 & 2.50 & 4784 & 2.48 & $-$0.35 & 4850 & $-$0.15 & 1.33 & 4.5 & 1.0 &  $-$32.2 & RC\\
 5701829 & 4850 & 3.23 & 1.03 & $-$0.20 & 4693 & 3.08 & 4623 & 4.63 & $-$0.48 & 4880 & $-$0.32 & 1.11 & 3.7 & 1.0 &  $-$20.5 & $-$ \\
 5709564 & 4745 & 2.33 & 1.50 & $-$0.27 & 4784 & 2.37 & 4752 & 2.52 & $-$0.06 & 4745 & $-$0.25 & 1.50 & 1.0 & 4.3 & $-$104.9 & RC\\
 5779724 & 4300 & 1.35 & 1.50 & $-$0.44 & $-$ & 1.68 & 4423 & 1.79 & $-$0.04 & 4240 & $-$0.14 & 1.50 & 3.0 & 3.2 &  $-$53.6 & $-$ \\
 5792581 & 4950 & 2.90 & 0.90 & $-$0.17 & $-$ & 2.82 & 4871 & 3.81 & $-$0.74 & 4980 & $-$0.22 & 0.90 & 2.4 & 2.4 &    5.2 & $-$ \\
 5795626 & 5100 & 2.77 & 1.50 & $-$0.69 & 4713 & 2.53 & 4990 & 2.70 & $-$1.01 & 5100 & $-$0.80 & 1.50 & 3.1 & 2.9 &  $-$89.1 & RC \\
 5859492 & 4770 & 2.55 & 1.25 &  0.38 & 4814 & 2.49 & 4632 & 2.01 &  0.37 & 4800 &  0.19 & 1.35 & 2.0 & 1.9 &  $-$59.4 & $-$ \\
 5866965 & 4155 & 1.75 & 1.30 & $-$0.24 & $-$ & 1.34 & 4163 & 1.93 & $-$1.02 & 4155 & $-$0.52 & 1.30 & 3.2 & 3.1 &  $-$64.9 & $-$ \\
 5900096 & 4580 & 2.35 & 1.50 &  0.30 & 4646 & 2.44 & 4539 & 1.95 &  0.54 & 4580 &  0.33 & 1.50 & 2.7 & 2.5 &  $-$44.7 & RC\\
 6101376 & 5100 & 2.59 & 1.48 & $-$0.07 & 5139 & 2.59 & $-$ & $-$ & $-$ & 5100 & $-$0.05 & 1.48 & 7.8 & 0.6 &  $-$19.3 & $-$ \\
 6125893 & 4280 & 1.60 & 1.50 &  0.13 & 4156 & 1.79 & 4346 & 1.92 &  0.27 & 4260 &  0.29 & 1.50 & 3.0 & 2.8 &  $-$35.9 & $-$ \\
 6465075 & 4820 & 2.65 & 1.35 & $-$0.35 & 4818 & 2.87 & 4667 & 3.27 & $-$1.18 & 4770 & $-$0.18 & 1.35 & 1.7 & 2.6 &   $-$4.0 & $-$ \\
 6547007 & 4785 & 2.50 & 1.40 & $-$0.67 & 4488 & 2.50 & 4757 & 2.84 & $-$0.59 & 4785 & $-$0.64 & 1.40 & 2.8 & 3.3 &  $-$94.1 & $-$ \\
 6579998 & 5050 & 2.40 & 1.50 & $-$0.73 & 4853 & 2.45 & 4969 & 2.89 & $-$0.69 & 5070 & $-$0.69 & 1.50 & 3.2 & 3.6 &  $-$43.4 & RC\\
 6680734 & 4580 & 2.15 & 1.50 & $-$0.43 & 4393 & 2.17 & 4597 & 2.57 & $-$0.53 & 4580 & $-$0.38 & 1.50 & 1.7 & 1.6 &   12.6 & $-$ \\
 6690139 & 5050 & 3.05 & 1.25 & $-$0.13 & 4800 & 3.00 & 4950 & 3.22 & $-$0.60 & 5020 & $-$0.13 & 1.25 & 3.1 & 3.1 &   13.5 & RGB\\
 6696436 & 4680 & 2.50 & 1.30 & $-$0.20 & 4469 & 2.33 & 4683 & 2.03 & $-$0.43 & 4630 & $-$0.26 & 1.30 & 3.1 & 3.0 &  $-$13.9 & $-$ \\
 6837256 & 4850 & 2.60 & 1.35 & $-$0.63 & 4576 & 2.48 & 4728 & 2.62 & $-$0.62 & 4850 & $-$0.65 & 1.35 & 1.7 & 1.4 &    0.9 & $-$ \\
 7006979 & 4870 & 2.37 & 1.40 & $-$0.25 & 4655 & 2.46 & 4891 & 2.21 & $-$0.01 & 4870 & $-$0.19 & 1.40 & 1.2 & 4.0 &  $-$57.3 & RC\\
 7340724 & 4879 & 3.05 & 1.14 &  0.05 & 4845 & 3.05 & 4867 & 2.34 & $-$0.11 & 4879 &  0.04 & 1.14 & 3.9 & 1.0 &   $-$8.2 & RGB\\
 7366121 & 4910 & 2.69 & 1.41 & $-$0.05 & 4872 & 2.59 & $-$ & $-$ & $-$ & 4910 & $-$0.10 & 1.41 & 4.2 & 1.0 &  $-$26.0 & RC \\
 7693833 & 4900 & 1.95 & 1.70 & $-$2.45 & 4800 & 2.46 & 4848 & 3.16 & $-$1.45 & 4880 & $-$2.23 & 1.70 & 2.3 & 2.2 &  $-$14.3 & RGB \\
 7812552 & 5070 & 3.50 & 0.95 & $-$0.44 & 4901 & 3.26 & 4930 & 3.35 & $-$0.87 & 5070 & $-$0.59 & 1.05 & 2.5 & 1.9 &   16.0 & $-$ \\
 7909976 & 5091 & 2.80 & 1.37 &  0.01 & 4934 & 2.80 & $-$ & $-$ & $-$ & 5091 &  0.00 & 1.37 & 4.5 & 0.6 &  $-$17.8 & RGB\\
 8017159 & 4625 & 1.11 & 1.70 & $-$2.05 & $-$ & 1.38 & 4634 & 2.45 & $-$1.07 & 4625 & $-$1.95 & 1.70 & 3.0 & 3.0 & $-$375.4 & $-$ \\
 8210100 & 4692 & 2.57 & 1.30 &  0.21 & $-$ & 2.53 & 4685 & 2.36 &  0.45 & 4692 &  0.20 & 1.30 & 4.8 & 1.0 &  $-$29.5 & RC\\
 8211551 & 4822 & 2.48 & 1.35 & $-$0.20 & 4905 & 2.48 & 4658 & 2.36 &  0.45 & 4822 & $-$0.20 & 1.35 & 4.5 & 1.0 &  $-$15.0 & $-$ \\
 8476245 & 4865 & 1.86 & 1.75 & $-$1.33 & 4372 & 1.96 & 4817 & 2.76 & $-$1.20 & 4865 & $-$1.28 & 1.70 & 3.0 & 3.0 & $-$129.2 & $-$ \\
 8491147 & 5020 & 2.70 & 1.40 & $-$0.39 & $-$ & 2.49 & 4770 & 3.34 & $-$1.03 & 5050 & $-$0.51 & 1.40 & 2.7 & 2.1 &    7.1 & $-$ \\
 8493969 & 4830 & 3.00 & 1.00 & $-$0.16 & 4652 & 2.97 & 4655 & 3.30 & $-$0.94 & 4830 & $-$0.20 & 1.00 & 1.6 & 1.3 &   $-$9.5 & $-$ \\
 8508931 & 5082 & 3.01 & 1.37 &  0.20 & 4847 & 2.71 & $-$ & $-$ & $-$ & 5082 & $-$0.02 & 1.37 & 4.6 & 3.2 &  $-$43.7 & SC \\
 8547390 & 4810 & 2.40 & 1.38 &  0.11 & 4804 & 2.59 & 4643 & 4.61 &  0.12 & 4780 &  0.24 & 1.38 & 5.2 & 1.0 &  $-$41.0 & RC\\
 8813946 & 4940 & 2.81 & 1.30 &  0.14 & 4929 & 2.81 & $-$ & $-$ & $-$ & 4940 &  0.14 & 1.30 & 4.8 & 1.0 &  $-$27.5 & RC\\
 8873797 & 4500 & 1.80 & 1.75 & $-$0.01 & 4793 & 2.41 & 4747 & 2.47 &  0.45 & 4500 &  0.32 & 1.75 & 3.5 & 3.4 &  $-$28.6 & RC \\
 9161068 & 5120 & 3.30 & 1.10 & $-$0.29 & 4960 & 3.26 & 4986 & 3.12 & $-$0.64 & 5120 & $-$0.29 & 1.10 & 2.1 & 2.6 &   21.2 & $-$ \\
 9288026 & 5050 & 2.50 & 1.25 & $-$0.35 & $-$ & 2.42 & 4855 & 3.41 & $-$0.74 & 5050 & $-$0.36 & 1.25 & 2.7 & 2.3 &   48.7 & $-$ \\
 9474021 & 4080 & 1.15 & 1.50 & $-$0.45 & $-$ & 1.20 & 4118 & 2.36 &  0.37 & 4080 & $-$0.47 & 1.50 & 2.8 & 2.8 & $-$124.0 & $-$ \\
 9532030 & 4450 & 1.91 & 1.26 &  0.02 & 4574 & 2.16 & 4408 & 4.59 &  0.33 & 4400 &  0.23 & 1.26 & 5.5 & 1.0 &  $-$13.5 & $-$ \\
 9574235 & 4380 & 0.95 & 2.25 & $-$1.30 & $-$ & $-$ & 4334 & 1.74 & $-$1.42 & $-$ & $-$ & $-$ & 3.1 & 3.8 &   $-$2.5 & $-$ \\
 9705687 & 5100 & 2.80 & 1.40 & $-$0.29 & $-$ & 2.79 & 5134 & 2.69 & $-$0.25 & 5100 & $-$0.27 & 1.40 & 3.5 & 1.5 &    4.7 & RC \\
10186608 & 4725 & 2.50 & 1.40 &  0.03 & $-$ & 2.43 & 4850 & 2.40 &  0.21 & 4725 &  0.00 & 1.40 & 2.8 & 2.6 &  $-$11.1 & RC \\
10323222 & 4676 & 2.75 & 1.28 &  0.17 & 4571 & 2.60 & $-$ & $-$ & $-$ & 4706 &  0.06 & 1.28 & 4.7 & 1.0 &  $-$22.7 & RGB \\
10403036 & 4485 & 1.90 & 1.35 & $-$0.59 & 4241 & 1.92 & 4388 & 2.21 & $-$1.39 & 4505 & $-$0.61 & 1.35 & 4.5 & 2.0 & $-$124.9 & $-$ \\
10404994 & 4855 & 2.55 & 1.33 & $-$0.02 & 4706 & 2.54 & $-$ & $-$ & $-$ & 4855 & $-$0.05 & 1.33 & 2.0 & 3.2 &   $-$0.7 & RC \\
10426854 & 4955 & 2.38 & 1.45 & $-$0.39 & $-$ & 2.50 & 4731 & 2.57 & $-$1.03 & 4955 & $-$0.37 & 1.45 & 4.5 & 1.5 &  $-$45.6 & RC \\
10649021 & 3960 & 0.80 & 1.45 & $-$0.30 & $-$ & 1.18 & 4083 & 1.79 &  0.54 & 3790 &  0.37 & 1.35 & 3.6 & 3.5 &  $-$39.1 & $-$ \\
10716853 & 4898 & 2.62 & 1.38 & $-$0.09 & 4761 & 2.62 & $-$ & $-$ & $-$ & 4898 & $-$0.10 & 1.38 & 4.3 & 1.0 &    2.3 & RC \\
11045542 & 4400 & 1.40 & 1.50 & $-$0.59 & $-$ & 1.75 & 4425 & 2.12 & $-$0.85 & 4450 & $-$0.51 & 1.50 & 2.4 & 2.2 &   17.3 & $-$ \\
11342694 & 4575 & 2.60 & 1.18 &  0.26 & 4758 & 2.82 & 4603 & 2.65 &  0.50 & 4575 &  0.38 & 1.18 & 2.0 & 3.5 &  $-$19.9 & $-$ \\
11444313 & 4750 & 2.40 & 1.50 & $-$0.02 & 4822 & 2.46 & 4888 & 2.47 &  0.17 & 4750 & $-$0.01 & 1.50 & 2.5 & 2.2 &  $-$17.8 & RC \\
11569659 & 4890 & 2.45 & 1.50 & $-$0.27 & 4680 & 2.43 & 5036 & 2.36 &  0.06 & 4890 & $-$0.27 & 1.50 & 3.1 & 3.1 &  $-$19.2 & RC \\
11657684 & 4840 & 2.00 & 1.45 & $-$0.32 & $-$ & 2.44 & 5066 & 2.60 &  0.25 & 4840 & $-$0.09 & 1.45 & 3.4 & 3.0 &   14.0 & RC \\
11674677 & 4973 & 2.49 & 1.36 & $-$0.18 & 4830 & 2.49 & $-$ & $-$ & $-$ & 4973 & $-$0.18 & 1.36 & 4.7 & 1.0 &    6.3 & RC \\
12455203 & 5080 & 2.84 & 1.30 &  0.04 & 4760 & 2.84 & 4983 & 3.14 & $-$0.12 & 5080 &  0.03 & 1.30 & 4.6 & 1.0 &  $-$10.8 & $-$  \\
12884274 & 4770 & 2.60 & 1.43 &  0.17 & 4946 & 2.60 & $-$ & $-$ & $-$ & 4770 &  0.18 & 1.43 & 4.9 & 3.2 &  $-$34.6 & RC \\
\end{longtable}
\normalsize
\end{landscape}
\twocolumn
\onecolumn
           \begin{landscape}
           \small
           \begin{longtable}{rrrrrrrrrrrrrrrrr}
           \caption{\label{abundseis} Elemental abundances for all targets where an asteroseismic \logg\ could be determined. Uncertainties are 0.5 dex for [C/H] and [O/H], 0.15 dex for [Fe/H] and 0.2 dex for the remaining elements. KIC9574235 has abundances based on the spectroscopic \logg.}\\
           \hline\hline
           KIC-ID  & [C/H] & [O/H] & [Si/H]& [Ca/H]& [Sc/H]& [Ti-I/H] & [Ti-II/H] & [V/H] & [Cr-I/H] & [Cr-II/H] & [Mn/H] & [Fe-I/H] & [Fe-II/H] & [Co/H] & [Ni/H] & [Y/H] \\ 
           \hline
           \endfirsthead
           \caption{Continued.}\\
           \hline\hline
           KIC-ID  & [C/H] & [O/H] & [Si/H]& [Ca/H]& [Sc/H]& [Ti-I/H] & [Ti-II/H] & [V/H] & [Cr-I/H] & [Cr-II/H] & [Mn/H] & [Fe-I/H] & [Fe-II/H] & [Co/H] & [Ni/H] & [Y/H] \\ 
           \hline
           \endhead
           \hline
					 \endfoot
					 \hline\hline
					 \endlastfoot	
					 1726211 & $-$ & $-$ & $-$0.34 & $-$0.45 & $-$0.36 & $-$0.36 & $-$ & $-$0.42 & $-$0.63 & $-$ & $-$ & $-$0.66 & $-$0.67 & $-$ & $-$0.60 & $-$   \\
           2425631 & $-$ & $-$ &  0.02 & $-$0.07 & $-$ & $-$0.15 & $-$0.23 &  0.00 & $-$0.26 & $-$ & $-$ & $-$0.12 & $-$0.03 & $-$0.25 & $-$0.17 & $-$   \\
           2714397 & $-$ & $-$ & $-$0.26 & $-$0.02 & $-$0.35 & $-$0.07 & $-$ & $-$0.12 & $-$0.44 & $-$ & $-$ & $-$0.42 & $-$0.59 & $-$ & $-$0.37 & $-$   \\
           3429205 & $-$ & $-$ &  0.04 & $-$ & $-$0.02 &  0.04 & $-$0.14 &  0.12 &  0.03 & $-$ & $-$ &  0.02 & $-$0.11 &  0.14 &  0.05 & $-$0.01   \\
           3430868 &  0.11 &  0.02 &  0.06 &  0.08 & $-$ &  0.02 &  0.02 &  0.10 &  0.00 & $-$0.01 & $-$ &  0.03 & $-$0.06 &  0.08 & $-$0.04 &  0.02   \\
           3644958 & $-$ & $-$ & $-$0.06 & $-$ & $-$ & $-$0.21 & $-$ & $-$0.17 & $-$0.26 & $-$0.21 & $-$ & $-$0.21 & $-$0.23 & $-$0.11 & $-$0.23 & $-$0.22   \\
           3744043 & $-$ & $-$ & $-$0.22 & $-$0.30 & $-$0.21 & $-$0.29 & $-$0.16 & $-$0.27 & $-$0.39 & $-$ & $-$0.38 & $-$0.32 & $-$0.31 & $-$0.18 & $-$0.28 & $-$0.32   \\
           3748585 & $-$ & $-$ &  0.33 &  0.09 & $-$ &  0.06 &  0.03 &  0.32 &  0.09 & $-$ & $-$ &  0.17 &  0.25 &  0.34 &  0.14 & $-$0.04   \\
           3748691 & $-$ & $-$ &  0.23 &  0.04 & $-$0.01 &  0.01 & $-$ &  0.04 & $-$0.08 & $-$ & $-$ &  0.11 &  0.13 &  0.21 &  0.06 &  0.02   \\
           3860139 & $-$ & $-$ &  0.38 &  0.22 & $-$ &  0.04 & $-$ &  0.46 & $-$0.07 & $-$ & $-$ &  0.12 &  0.10 &  0.30 &  0.11 & $-$   \\
           3936921 & $-$ & $-$ &  0.36 & $-$ &  0.02 &  0.02 & $-$ &  0.16 & $-$0.10 & $-$ & $-$ &  0.03 &  0.09 &  0.32 &  0.06 & $-$   \\
           3955590 & $-$ & $-$ &  0.27 & $-$ & $-$ &  0.36 & $-$ &  0.72 &  0.26 & $-$ & $-$ &  0.23 & $-$0.16 &  0.42 &  0.23 &  0.03   \\
           4070746 & $-$ & $-$ & $-$0.01 & $-$ & $-$0.10 & $-$0.09 & $-$ &  0.01 & $-$0.15 & $-$ & $-$ & $-$0.04 & $-$0.17 &  0.00 & $-$0.12 & $-$   \\
           4072740 & $-$ &  0.30 &  0.30 &  0.19 &  0.19 &  0.08 &  0.03 &  0.14 &  0.07 &  0.09 & $-$ &  0.16 &  0.23 &  0.38 &  0.22 &  0.07   \\
           4157282 & $-$ & $-$ &  0.27 & $-$0.14 & $-$ & $-$0.16 & $-$ &  0.07 & $-$0.34 & $-$ & $-$ &  0.02 &  0.24 & $-$ & $-$0.01 & $-$   \\
           4177025 & $-$ & $-$ &  0.04 & $-$0.37 & $-$ & $-$0.28 & $-$ & $-$0.04 & $-$0.64 & $-$ & $-$ & $-$0.36 & $-$0.24 & $-$ & $-$0.27 & $-$   \\
           4262505 & $-$ & $-$ & $-$0.04 & $-$0.16 & $-$ & $-$0.15 & $-$ & $-$0.06 & $-$0.27 & $-$ & $-$ & $-$0.18 & $-$0.18 & $-$0.06 & $-$0.19 & $-$0.19   \\
           4283484 & $-$ & $-$ & $-$0.48 & $-$ & $-$ & $-$0.62 & $-$ & $-$0.69 & $-$0.97 & $-$ & $-$ & $-$0.79 & $-$0.78 & $-$0.63 & $-$0.81 & $-$0.81   \\
           4480358 & $-$ & $-$ & $-$0.50 & $-$0.64 & $-$0.56 & $-$0.48 & $-$0.76 & $-$0.56 & $-$0.86 & $-$ & $-$ & $-$0.85 & $-$0.96 & $-$0.55 & $-$0.83 & $-$0.91   \\
           4659706 & $-$ & $-$ &  0.78 & $-$ & $-$ &  0.36 & $-$ &  0.63 &  0.46 & $-$ & $-$ &  0.47 &  0.62 &  0.72 &  0.59 & $-$   \\
           5113061 & $-$ & $-$ &  0.32 & $-$ & $-$ &  0.19 &  0.05 &  0.45 &  0.08 & $-$ & $-$ &  0.11 &  0.01 & $-$ &  0.02 & $-$   \\
           5113910 & $-$ & $-$ & $-$0.10 & $-$0.30 & $-$ & $-$0.31 & $-$0.34 & $-$0.10 & $-$0.50 & $-$ & $-$ & $-$0.32 & $-$0.31 & $-$0.03 & $-$0.38 & $-$   \\
           5284127 & $-$ & $-$ &  0.60 & $-$ & $-$ &  0.50 & $-$ &  0.72 &  0.38 & $-$ & $-$ &  0.45 &  0.45 &  0.75 &  0.51 &  0.33   \\
           5511423 & $-$ & $-$ & $-$0.64 & $-$0.96 & $-$1.36 & $-$1.03 & $-$ & $-$1.15 & $-$ & $-$ & $-$ & $-$1.10 & $-$0.96 & $-$ & $-$1.17 & $-$  \\
           5524720 & $-$ & $-$ &  0.44 & $-$ & $-$ &  0.19 & $-$ &  0.34 &  0.30 & $-$ & $-$ &  0.24 &  0.38 &  0.59 &  0.26 & $-$   \\
           5612549 & $-$ & $-$ & $-$0.16 & $-$0.45 & $-$ & $-$0.43 & $-$ & $-$0.43 & $-$0.49 & $-$ & $-$ & $-$0.35 & $-$0.34 & $-$0.25 & $-$0.34 & $-$0.37   \\
           5698156 & $-$ & $-$ & $-$0.82 & $-$1.00 & $-$1.35 & $-$1.03 & $-$0.96 & $-$1.20 & $-$1.42 & $-$ & $-$ & $-$1.30 & $-$1.30 & $-$1.16 & $-$1.23 & $-$1.11   \\
           5700368 &  0.11 &  0.12 & $-$0.05 & $-$0.13 & $-$0.13 & $-$0.21 & $-$0.10 & $-$0.16 & $-$0.26 & $-$0.24 & $-$0.26 & $-$0.15 & $-$0.15 & $-$0.10 & $-$0.19 & $-$0.15   \\
           5701829 & $-$0.21 & $-$0.04 & $-$0.04 & $-$0.06 & $-$0.06 &  0.00 & $-$0.05 &  0.11 & $-$0.21 & $-$0.27 & $-$ & $-$0.23 & $-$0.33 & $-$0.06 & $-$0.19 & $-$0.33   \\
           5709564 & $-$ & $-$ &  0.07 & $-$0.13 & $-$0.22 & $-$0.07 & $-$ & $-$0.05 & $-$0.32 & $-$ & $-$0.50 & $-$0.26 & $-$0.26 &  0.02 & $-$0.19 & $-$   \\
           5779724 & $-$ & $-$ &  0.01 & $-$0.27 & $-$ & $-$0.37 & $-$ & $-$0.37 & $-$0.52 & $-$ & $-$ & $-$0.35 & $-$0.14 & $-$0.01 & $-$0.29 & $-$   \\
           5792581 & $-$ & $-$ & $-$0.16 & $-$0.06 & $-$0.13 & $-$0.13 & $-$0.22 & $-$0.01 & $-$0.18 & $-$ & $-$ & $-$0.17 & $-$0.22 & $-$0.11 & $-$0.24 & $-$0.16   \\
           5795626 & $-$ & $-$ & $-$0.41 & $-$0.47 & $-$0.51 & $-$0.39 & $-$0.44 & $-$0.48 & $-$0.74 & $-$0.67 & $-$1.00 & $-$0.71 & $-$0.78 & $-$0.55 & $-$0.71 & $-$0.76   \\
           5859492 & $-$ & $-$ &  0.47 &  0.26 & $-$ &  0.41 &  0.10 &  0.66 &  0.35 & $-$ & $-$ &  0.30 &  0.19 &  0.54 &  0.30 & $-$   \\
           5866965 & $-$ & $-$ &  $-$0.05 & $-$0.09 & $-$0.47 &  0.03 & $-$ &  0.17 & $-$0.46 & $-$ & $-$ & $-$0.36 & $-$0.52 &  0.36 & $-$0.39 & $-$   \\
           5900096 & $-$ & $-$ &  0.60 & $-$ & $-$ &  0.39 & $-$ &  0.80 &  0.29 & $-$ & $-$ &  0.32 &  0.33 & $-$ &  0.41 &  0.04   \\
           6101376 &  0.06 &  0.10 &  0.01 & $-$0.08 & $-$0.01 & $-$0.18 & $-$0.05 & $-$0.06 & $-$0.20 & $-$0.26 &  0.00 & $-$0.08 & $-$0.07 & $-$0.17 & $-$0.18 & $-$0.09   \\
           6125893 & $-$ & $-$ &  0.43 & $-$ & $-$ &  0.09 & $-$ & $-$ &  0.04 & $-$ & $-$ &  0.22 &  0.26 & $-$ &  0.18 & $-$   \\
           6465075 & $-$ & $-$ & $-$0.04 & $-$0.40 & $-$ & $-$0.43 & $-$ & $-$0.48 & $-$0.58 & $-$ & $-$ & $-$0.33 & $-$0.18 & $-$0.23 & $-$0.34 & $-$   \\
           6547007 & $-$ & $-$ & $-$0.39 & $-$0.62 & $-$0.58 & $-$0.59 & $-$0.38 & $-$0.60 & $-$0.76 & $-$ & $-$0.72 & $-$0.67 & $-$0.67 & $-$0.57 & $-$0.63 & $-$0.69   \\
           6579998 & $-$ & $-$ & $-$0.43 & $-$0.61 & $-$0.56 & $-$0.45 & $-$0.26 & $-$0.56 & $-$0.80 & $-$0.77 & $-$1.20 & $-$0.71 & $-$0.70 & $-$0.54 & $-$0.67 & $-$   \\
           6680734 & $-$ & $-$ & $-$0.11 & $-$0.42 & $-$0.40 & $-$0.43 & $-$ & $-$0.33 & $-$0.56 & $-$ & $-$ & $-$0.42 & $-$0.40 & $-$0.26 & $-$0.43 & $-$0.59   \\
           6690139 & $-$ & $-$ & $-$0.04 & $-$0.05 &  0.11 & $-$0.08 & $-$ & $-$0.18 & $-$0.31 & $-$ &  0.09 & $-$0.15 & $-$0.14 &  0.01 & $-$0.14 & $-$   \\
           6696436 & $-$ & $-$ & $-$0.13 & $-$0.25 & $-$ & $-$0.13 & $-$ & $-$0.03 & $-$0.34 & $-$ & $-$ & $-$0.23 & $-$0.24 & $-$0.01 & $-$0.22 & $-$   \\
           6837256 & $-$ & $-$ & $-$0.34 & $-$0.34 & $-$ & $-$0.34 & $-$0.44 & $-$0.40 & $-$0.66 & $-$ & $-$ & $-$0.64 & $-$0.65 & $-$0.43 & $-$0.60 & $-$0.60   \\
           7006979 & $-$ & $-$ & $-$0.08 & $-$0.33 & $-$0.23 & $-$0.24 & $-$ & $-$0.27 & $-$0.40 & $-$ & $-$ & $-$0.27 & $-$0.19 & $-$0.18 & $-$0.24 & $-$   \\
           7340724 & $-$ &  0.12 &  0.16 &  0.08 &  0.07 &  0.03 &  0.04 &  0.25 &  0.00 & $-$0.06 &  0.30 &  0.05 &  0.05 &  0.18 &  0.05 &  0.00   \\
           7366121 & $-$ &  0.00 &  0.04 & $-$0.02 & $-$0.10 & $-$0.07 & $-$0.06 &  0.08 & $-$0.12 & $-$0.19 & $-$ & $-$0.05 & $-$0.10 &  0.02 & $-$0.06 & $-$0.07   \\
           7693833 & $-$ & $-$ & $-$ & $-$2.09 & $-$2.27 & $-$2.42 & $-$ & $-$ & $-$2.99 & $-$ & $-$ & $-$2.48 & $-$2.23 & $-$ & $-$2.29 & $-$2.21   \\
           7812552 & $-$ & $-$ & $-$0.28 & $-$0.22 & $-$0.27 & $-$0.17 & $-$0.25 & $-$0.22 & $-$0.40 & $-$ & $-$ & $-$0.46 & $-$0.59 & $-$0.24 & $-$0.45 & $-$0.52   \\
           7909976 &  0.13 & $-$0.02 &  0.06 & $-$ & $-$ & $-$0.03 & $-$0.01 &  0.03 & $-$0.06 & $-$0.10 & $-$ &  0.01 &  0.01 & $-$0.01 & $-$0.07 &  0.09   \\
           8017159 & $-$ & $-$ & $-$1.37 & $-$1.75 & $-$2.52 & $-$2.03 & $-$ & $-$ & $-$2.32 & $-$ & $-$ & $-$1.99 & $-$1.95 & $-$ & $-$2.11 & $-$   \\
           8210100 & $-$ &  0.26 &  0.39 &  0.13 & $-$ &  0.12 &  0.17 &  0.37 &  0.08 &  0.23 & $-$ &  0.21 &  0.22 &  0.41 &  0.23 &  0.19   \\
           8211551 & $-$0.59 & $-$0.04 & $-$0.06 & $-$0.17 & $-$0.12 & $-$0.21 & $-$0.14 & $-$0.15 & $-$0.37 & $-$0.31 & $-$0.05 & $-$0.20 & $-$0.20 & $-$0.07 & $-$0.18 & $-$0.28   \\
           8476245 & $-$ & $-$ & $-$0.99 & $-$0.93 & $-$1.22 & $-$1.17 & $-$ & $-$ & $-$1.48 & $-$ & $-$1.82 & $-$1.32 & $-$1.30 & $-$ & $-$1.34 & $-$   \\
           8491147 & $-$ & $-$ & $-$0.29 & $-$0.17 & $-$ & $-$0.28 & $-$0.37 & $-$0.28 & $-$0.39 & $-$ & $-$ & $-$0.41 & $-$0.51 & $-$0.31 & $-$0.42 & $-$0.51   \\
           8493969 & $-$ & $-$ &  0.10 & $-$0.02 & $-$ & $-$0.02 & $-$ &  0.04 & $-$0.16 & $-$ & $-$ & $-$0.17 & $-$0.19 &  0.06 & $-$0.18 & $-$   \\
           8508931 & $-$ & $-$0.10 &  0.19 &  0.28 &  0.24 &  0.16 &  0.03 &  0.31 &  0.13 & $-$0.05 & $-$ &  0.16 & $-$0.02 &  0.19 &  0.12 &  0.16   \\
           8547390 & $-$ &  0.27 &  0.25 &  0.12 & $-$ &  0.05 &  0.10 &  0.14 &  0.02 & $-$ & $-$ &  0.14 &  0.25 &  0.26 &  0.13 &  0.17   \\
           8813946 & $-$ &  0.32 &  0.24 &  0.15 &  0.19 &  0.05 &  0.15 &  0.19 &  0.07 &  0.10 & $-$ &  0.14 &  0.14 &  0.28 &  0.14 &  0.17   \\
           8873797 & $-$ & $-$ &  0.59 & $-$ & $-$ &  0.09 &  0.30 & $-$0.08 & $-$ & $-$ & $-$ &  0.14 &  0.32 &  0.44 &  0.14 & $-$   \\
           9161068 & $-$ & $-$ & $-$0.28 & $-$ & $-$0.25 & $-$0.22 & $-$0.33 & $-$0.24 & $-$0.33 & $-$ & $-$ & $-$0.29 & $-$0.29 & $-$0.26 & $-$0.30 & $-$0.37   \\
           9288026 & $-$ & $-$ & $-$0.29 & $-$0.35 & $-$ & $-$0.32 & $-$0.25 & $-$0.36 & $-$0.45 & $-$0.40 & $-$0.35 & $-$0.36 & $-$0.36 & $-$0.27 & $-$0.40 & $-$0.45   \\
           9474021 & $-$ & $-$ &  0.09 & $-$0.32 & $-$ & $-$0.07 & $-$0.35 & $-$0.09 & $-$0.52 & $-$ & $-$ & $-$0.47 & $-$0.47 &  0.02 & $-$0.41 & $-$   \\
           9532030 & $-$ &  0.43 &  0.26 &  0.03 &  0.09 & $-$0.07 &  0.00 &  0.12 & $-$0.11 & $-$ &  0.13 &  0.09 &  0.23 &  0.23 &  0.05 &  0.14   \\
					 9574235 & $-$ & $-$ & $-$0.96 & $-$1.12 & $-$ & $-$1.10 & $-$1.41 & $-$1.22 & $-$1.61 & $-$1.89 & $-$ &  $-$1.30 & $-$1.27 & $-$1.23 & $-$1.45 & $-$   \\
           9705687 & $-$ & $-$ & $-$0.14 & $-$ & $-$0.16 & $-$0.30 & $-$0.28 & $-$0.34 & $-$0.35 & $-$ & $-$ & $-$0.29 & $-$0.27 & $-$0.25 & $-$0.35 & $-$   \\
          10186608 & $-$ & $-$ &  0.20 & $-$0.06 & $-$0.03 & $-$0.04 & $-$ &  0.15 & $-$0.07 & $-$ & $-$ &  0.02 &  0.00 &  0.20 &  0.05 & $-$0.15   \\
          10323222 &  1.04 &  0.39 &  0.25 &  0.27 &  0.20 &  0.19 &  0.08 &  0.40 &  0.10 & $-$ & $-$ &  0.15 &  0.05 &  0.40 &  0.19 & $-$0.02   \\
          10403036 & $-$ & $-$ & $-$0.21 & $-$ & $-$0.40 & $-$0.18 & $-$ & $-$0.35 & $-$0.52 & $-$ & $-$ & $-$0.58 & $-$0.59 & $-$0.27 & $-$0.49 & $-$   \\
          10404994 &  0.15 &  0.11 &  0.10 & $-$0.03 & $-$0.02 & $-$0.05 & $-$0.03 &  0.06 & $-$0.11 & $-$0.21 & $-$ & $-$0.01 & $-$0.02 &  0.11 & $-$0.04 & $-$0.06   \\
          10426854 & $-$ & $-$ & $-$0.27 & $-$0.34 & $-$ & $-$0.42 & $-$ & $-$0.46 & $-$ & $-$ & $-$ & $-$0.42 & $-$0.37 & $-$0.30 & $-$0.42 & $-$0.31   \\
          10649021 & $-$ & $-$ &  0.40 & $-$ & $-$ & $-$0.14 &  0.26 &  0.11 & $-$0.26 & $-$ & $-$ & $-$0.04 &  0.37 &  0.31 & $-$0.03 & $-$   \\
          10716853 &  0.16 &  0.08 &  0.03 & $-$0.07 & $-$0.09 & $-$0.13 & $-$0.07 & $-$0.02 & $-$0.18 &  0.20 & $-$0.18 & $-$0.09 & $-$0.09 & $-$0.04 & $-$0.12 & $-$0.01   \\
          11045542 & $-$ & $-$ & $-$0.31 & $-$0.49 & $-$0.60 & $-$0.40 & $-$0.39 & $-$0.27 & $-$0.63 & $-$ & $-$ & $-$0.53 & $-$0.51 & $-$0.25 & $-$0.53 & $-$0.70   \\
          11342694 & $-$ & $-$ &  0.50 &  0.24 & $-$ &  0.28 &  0.34 &  0.68 &  0.18 & $-$ & $-$ &  0.34 &  0.36 &  1.01 &  0.40 & $-$   \\
          11444313 & $-$ & $-$ &  0.15 & $-$0.08 & $-$ & $-$0.06 & $-$0.10 &  0.11 & $-$0.14 & $-$ & $-$ & $-$0.02 & $-$0.02 &  0.18 & $-$0.05 & $-$0.12   \\
          11569659 & $-$ & $-$ & $-$0.16 & $-$ & $-$ & $-$0.22 & $-$0.26 & $-$0.20 & $-$0.37 & $-$ & $-$ & $-$0.27 & $-$0.27 & $-$0.18 & $-$0.27 & $-$   \\
          11657684 & $-$ & $-$ & $-$0.03 & $-$0.23 & $-$0.09 & $-$0.31 & $-$0.19 & $-$0.29 & $-$0.39 & $-$ & $-$ & $-$0.28 & $-$0.09 & $-$0.12 & $-$0.25 &  0.17   \\
          11674677 & $-$0.11 & $-$ &  0.08 & $-$0.17 & $-$0.12 & $-$0.20 & $-$0.15 & $-$0.14 & $-$0.26 & $-$0.35 & $-$0.15 & $-$0.18 & $-$0.18 & $-$0.13 & $-$0.26 & $-$0.16   \\
          12455203 &  0.00 &  0.02 &  0.08 &  0.10 &  0.02 &  0.00 &  0.10 &  0.09 & $-$0.02 & $-$0.05 & $-$ &  0.05 &  0.04 &  0.12 & $-$0.01 &  0.08   \\
          12884274 &  0.94 &  0.29 &  0.30 &  0.18 & $-$ &  0.06 &  0.18 &  0.34 &  0.06 & $-$0.01 & $-$ &  0.17 &  0.17 &  0.37 &  0.22 &  0.07   \\
          \end{longtable}
          \normalsize
          \end{landscape}
\twocolumn

% ++++++++++++++++++++++++++++++++++
% ++++++++++++++++++++++++++++++++++
% ++++++++++++++++++++++++++++++++++
\end{document}